\documentclass[aps,prl,reprint,amssymb,longbibliography]{revtex4-2}
\newcommand{\be}{\begin{equation}}
\newcommand{\ee}{\end{equation}}
\newcommand{\bea}{\begin{eqnarray}}
\newcommand{\eea}{\end{eqnarray}}
\newcommand{\ii}{\mathrm{i}}
\usepackage{graphicx}
\usepackage{amsmath}
\usepackage{color}
\usepackage{dcolumn}
\usepackage{bm}
\usepackage{hyperref}
\usepackage{braket}
\usepackage{enumerate}

\begin{document}

\title{Composite-Boson Ansatz for Fractional Quantum Hall Manifolds of Lattice Bosons at Generic Fillings $\nu<1/2$}
\author{Umut Can Turhan}
\email{can.turhan@bilkent.edu.tr}
\affiliation{Department of Physics, Bilkent University, Ankara, T\"urkiye}
\author{Ahmet Levent Suba\c s\i}
\email{alsubasi@itu.edu.tr}
\affiliation{Department of Physics, Istanbul Technical University, \.{I}stanbul, T\"urkiye}
\author{Rifat Onur Umucal\i lar}
\email{onur.umucalilar@msgsu.edu.tr}
\affiliation{Department of Physics, Mimar Sinan Fine Arts University, \.{I}stanbul, T\"urkiye}

\date{\today}

\begin{abstract}
Hofstadter systems provide a lattice route to fractional quantum Hall physics, but their low-energy manifolds often lack simple wave-function descriptions. Here we present a composite-boson ansatz for a broad family of finite-size lattice-split Laughlin-quasihole
manifolds in the low-flux Hofstadter--Bose--Hubbard model on a torus at generic fillings \(\nu<1/2\). Attaching two vortices to each boson yields reduced-flux orbitals from which we
construct a many-body trial basis. Its translation-resolved rank reproduces the
expected low-energy-manifold dimension and provides a composite-boson
interpretation of the established quasihole counting previously inferred from
generalized exclusion rules and thin-torus arguments. For smaller systems, diagonalization within the lowest-band-projected ansatz
span closely reproduces the exact projected spectrum, while variational Monte Carlo extends the
rank and energetic analysis to larger systems for which explicit subspace
construction becomes costly. To probe the fractional quantum Hall character of these manifolds, we
calculate their many-body Chern-numbers and show that localized
added-flux excitations exhibit quasihole-like behavior, with fractional density
depletion and an Aharonov--Bohm-free braiding phase that both track the effective
filling. Together, these results establish a microscopic composite-boson framework for organizing a broad family of sub-half-filled fractional quantum Hall
manifolds of lattice bosons.
\end{abstract}

\maketitle

{\it Introduction---} 
Studies of the fractional quantum Hall (FQH) effect~\cite{Tsui1982,Yoshioka2002} have shown that strongly interacting particles in two dimensions under a magnetic field form topological quantum liquids with fractionally charged excitations~\cite{Laughlin1983,dePicciotto1997,Saminadayar1997} and anyonic statistics~\cite{Leinaas1977,Arovas1984,Stern2008,Nayak2008,Bartolomei2020,Nakamura2020}. Its microscopic descriptions expose simple many-body organizing principles, notably Laughlin's wave function~\cite{Laughlin1983} and the composite-fermion construction~\cite{Jain2007,Hansson2017}. Despite important exact parent-Hamiltonian constructions~\cite{KapitMueller2010,GlasserCiracSierraNielsen2016}, transparent descriptions of generic lattice realizations remain challenging, particularly for Hofstadter~\cite{Sorensen2005,Hafezi2007,MollerCooper2009,MollerCooper2015} and fractional Chern-insulator (FCI) systems~\cite{RegnaultBernevig2011,LiuBergholtzFanLauchli2012,BernevigRegnault2012,LiuRepellinBernevigRegnault2013,SterdyniakRepellinBernevigRegnault2013,McGreevySwingleTran2012,Parameswaran2013}, where lattice structure and interactions modify continuum FQH physics and complicate its finite-size identification.

The Hofstadter--Bose--Hubbard (HBH) model provides a natural setting for
FQH-like states of interacting bosons in a topological band
~\cite{Sorensen2005,Hafezi2007,MollerCooper2009,
HeGrusdtKaufmanGreinerVishwanath2017,RepellinLeonardGoldman2020,
Leonard2023}. In exact diagonalization, these states appear as robust
quasidegenerate low-energy manifolds~\cite{Hafezi2007,MollerCooper2009},
whose topological character can be probed through many-body Chern
numbers~\cite{NiuThoulessWu1985,Hafezi2007,
KudoKariyadoHatsugai2017,ZeybekUmucalilar2022}. Topological entanglement entropy provides further evidence for Abelian
order at selected non-Laughlin fillings, notably for fermionic Hofstadter
states at \(\nu=2/5\) and \(3/7\)~\cite{AndrewsMohanNeupert2021}, while
spectral flow and entanglement spectra provide complementary diagnostics
in FCI models~\cite{RegnaultBernevig2011,BernevigRegnault2012}.
More broadly, quasiholes and their fractional properties have been studied
through local pinning, density measurements, and related probes in continuum
and lattice settings~\cite{Paredes2001,Raciunas2018,
Umucalilar2018Probe,UmucalilarMacalusoComparinCarusotto2018,
Macaluso2020,WangDongEckardt2022}, including explicit braiding in lattice
models~\cite{KapitGinspargMueller2011,LiuBhattRegnault2015,
JaworowskiRegnaultLiu2019}. To our knowledge, however, an explicit real-space braiding analysis has not
been reported for the generic sub-half-filled HBH manifolds considered here.

On the torus, many-body magnetic translations separate center-of-mass
degeneracy from the remaining quantum numbers~\cite{Haldane1985}, while
periodic Laughlin--Jastrow wave functions provide explicit continuum
states~\cite{HaldaneRezayi1985,GreiterSchnellsThomale2016}. In FCIs,
emergent many-body translation symmetries lead to a folding relation
between continuum FQH and lattice momentum
sectors~\cite{BernevigRegnault2012}. Quasihole counting likewise follows
from generalized Pauli principles~\cite{RegnaultBernevig2011}, Jack
polynomials~\cite{BernevigHaldane2008}, or thin-torus
arguments~\cite{SeidelYang2011,Hansson2017}. For the HBH model, a
previous study coauthored by one of us found a compact combinatorial
degeneracy formula, many-body Chern-number structure, and fractional
charge depletion for quasidegenerate manifolds below the bosonic
\(\nu=1/2\) parent state~\cite{ZeybekUmucalilar2022}. Taken together, however, these counting rules and diagnostics do not by
themselves provide a compact microscopic wave-function construction that
reorganizes the continuum quasihole space in terms of composite degrees of
freedom and accounts for the counting and internal structure of generic
lattice-split Laughlin-quasihole manifolds.

Composite-particle ideas are deeply rooted in FQH physics. Composite-boson Chern--Simons theory gives an effective-field-theory description of fractional Hall liquids~\cite{ZhangHanssonKivelson1989}, a distinct atom--vortex composite-boson picture was proposed for rotating condensates~\cite{WilkinGunn2000}, and nonlocal composite-boson order has been studied for the \(\nu=1/2\) HBH ground state~\cite{PauwEtAl2024}. While composite fermions successfully explain the Jain hierarchy~\cite{Jain2007,BergholtzHanssonHermannsKarlhede2007,Hansson2017}, efficient Landau-level projection schemes~\cite{JainKamilla1997IJMPB,JainKamilla1997PRB} require special treatment for torus wave functions~\cite{Hermanns2013,PuWuJain2017}. For rotating and Hofstadter bosons, one-vortex attachment was introduced early and developed into successful composite-fermion descriptions~\cite{CooperWilkin1999,RegnaultJolicoeur2003,RegnaultJolicoeur2004,ChangRegnaultJolicoeurJain2005,RegnaultChangJolicoeurJain2006,MollerCooper2009}. At \(\nu<1/2\), however, the ideal continuum lowest-Landau-level problem with contact interactions has a proliferating zero-energy Laughlin-quasihole sector, represented in planar geometry by the Laughlin--Jastrow factor multiplied by an arbitrary symmetric polynomial~\cite{ViefersHanssonReimann2000,MazaheriOrtizNussinovSeidel2015}. Recent work includes a real-space composite-boson framework and generalized
lattice-state constructions~\cite{YuZhu2026,JiWang2026}, variational
reconstructions of the full two- and three-quasihole manifolds in generalized
Kapit--Mueller and checkerboard models using position-localized parton
quasihole states and Gram-matrix analysis~\cite{LotricSimon2026}, and
single-anyon Bloch states and dispersions in ideal Chern
bands~\cite{IyerEtAl2026}.

Here we give an explicit reduced-flux torus organization of this established
Laughlin--quasihole space through an unpinned, statistics-preserving
two-vortex composite-boson (CB) ansatz built from continuum wave functions
sampled at lattice sites, and test how this space is realized in finite-size,
lattice-split manifolds of the nearest-neighbor low-flux HBH model at generic
fillings \(\nu<1/2\). For \(N\)
bosons pierced by \(N_\phi\) flux quanta, attaching two vortices to each boson
leaves \(N_d=N_\phi-2N\) effective orbitals. Naive occupation counting in this
reduced bosonic space is overcomplete, as in linearly dependent
composite-fermion trial spaces~\cite{MeyerLiaboeViefers2016}. Magnetic
translations organize the occupation patterns into cyclic orbits, reducing the
number of independent internal seeds, while the center-of-mass sectors supply
the complementary multiplicity. Together, these structures yield the physical
rank and reproduce the observed low-energy-manifold degeneracy. Overlap-matrix
ranks confirm this counting, while ansatz-subspace diagonalization~\cite{MandalJain2002,Jain2007,MeyerSreejithViefers2014} and
variational Monte Carlo (VMC)~\cite{Metropolis1953,FoulkesMitasNeedsRajagopal2001,NightingaleMelikAlaverdian2001,UmrigarFilippi2005,ToulouseUmrigar2007,BalramTokeWojsJain2015,WangGeraedtsRezayiHaldane2019} test
the resulting states' energetics. For the unpinned manifolds, we find a total many-body Chern number
\(C_{\rm MB}\) satisfying \(C_{\rm MB}/\mathcal D=\nu\), consistent with
their FQH character~\cite{NiuThoulessWu1985,KudoKariyadoHatsugai2017}.
With local pinning, an added-flux excitation develops a localized fractional
depletion \(Q=\nu_{\rm eff}\), while an Aharonov--Bohm-free braid yields a
phase consistent with \(2\nu_{\rm eff}\), supported by disorder and
asymmetric-pinning checks in a smaller system. Together, these results
establish a microscopic composite-boson organization of these generic
sub-half-filled manifolds and show that localized excitations within them
exhibit a coherent quasihole-like fractional response.

{\it The Model---}
We consider interacting bosons on a square lattice described by the HBH Hamiltonian~\cite{Langbein1969,Hofstadter1976,Sorensen2005,Hafezi2007,MollerCooper2009,
HeGrusdtKaufmanGreinerVishwanath2017,RepellinLeonardGoldman2020,
Leonard2023,Aidelsburger2013,Miyake2013}, supplemented, when needed, by local pinning potentials,
\begin{equation}
H(\lambda)=H_{\rm HBH}+H_{\rm pin}(\lambda),
\label{eq:Hamiltonian_full}
\end{equation}
with
\begin{equation}
H_{\rm HBH}
=
-t\sum_{\langle ij\rangle}
\left(
e^{i2\pi\phi_{ij}}c_i^\dagger c_j+\text{h.c.}
\right)
+
\frac{U}{2}\sum_i n_i(n_i-1),
\label{eq:Hamiltonian_HBH}
\end{equation}
and
\begin{equation}
H_{\rm pin}(\lambda)=\sum_{\alpha} V_\alpha(\lambda)\, n_{i_\alpha}.
\label{eq:Hamiltonian_pin}
\end{equation}
Here \(c_i^\dagger\) (\(c_i\)) creates (annihilates) a boson at site \(i\),
\(n_i=c_i^\dagger c_i\), \(t>0\) is the nearest-neighbor hopping amplitude,
and \(U>0\) the repulsive on-site interaction. \(H_{\rm pin}\) represents
local repulsive potentials at sites \(i_\alpha\): one or two static centers
localize quasiholes, while several time-dependent strengths
\(V_\alpha(\lambda)\) are varied adiabatically during braiding to transport
the density depletion along a prescribed path.

We work in the Landau gauge \({\bf A}=-By\,\hat{\bf x}\) corresponding to a magnetic field \(B\hat{\bf z}\) and define
\begin{equation}
\phi_{ij}=\frac{1}{\phi_0}
\int_{{\bf r}_i}^{{\bf r}_j}{\bf A}\cdot d{\bf r},
\label{eq:peierls_phase}
\end{equation}
where the integral follows the straight nearest-neighbor bond and
\(\phi_0=h/q_0\) is the flux quantum for effective charge \(q_0\). Thus
\(y\)-direction hoppings carry no Peierls phase, whereas \(x\)-direction
hoppings acquire a position-dependent phase. The dimensionless plaquette flux
is \(\phi=Ba_\ell^2/\phi_0>0\), with \(a_\ell\) the lattice spacing; the
counter-clockwise product of hopping phases around an elementary plaquette is
\(e^{-i2\pi\phi}\). The magnetic length is
\(\ell=a_\ell/\sqrt{2\pi\phi}\).

{\it Degeneracy counting and the composite boson ansatz---}
Below the bosonic \(\nu=1/2\) parent, the excess flux is
\(N_d=N_\phi-2N\). Consistent with generalized-Pauli-principle and thin-torus counting~\cite{RegnaultBernevig2011,SeidelYang2011}, our previous HBH analysis~\cite{ZeybekUmucalilar2022} obtained the torus degeneracy for \(N_d\) delocalized quasiholes as
\(\mathcal D=\binom{N_d+N-1}{N_d}N_\phi/N\). Its binomial prefactor equals
the numerical many-body Chern number of the manifold. Here we reorganize the
same formula algebraically for \(N_d>0\) as
\begin{equation}
\mathcal D
=
\binom{N_d+N-1}{N}\frac{N_\phi}{N_d}.
\label{eq:CB_counting_formula}
\end{equation}
The new binomial factor \(C_{\rm B}=\binom{N_d+N-1}{N}\) is simply the number of ways of placing \(N\) bosons in \(N_d\) single-particle orbitals.  This observation is the basic motivation for our composite-boson ansatz: after attaching two vortices to each physical boson, the low-energy manifold is organized by a reduced-flux bosonic Hilbert space, while magnetic translations and center-of-mass sectors supply the remaining factors in Eq.~\eqref{eq:CB_counting_formula}.

We now formulate this idea at the wave-function level for the unpinned
system; pinned quasiholes are used below as independent diagnostics.  On a
rectangular torus with periods \(L_1\) and \(\ii L_2\), we define
\(z_i=x_i+\ii y_i\), \(\tau=\ii L_2/L_1\), and \(Z=\sum_i z_i\).
Let \(\mathbf m=(m_1,\ldots,m_N)\) denote a list of reduced-flux orbital
labels whose multiplicities define the occupation pattern \(\mathbf n\).
Suppressing the real Landau-gauge Gaussian factors, we write
\begin{equation}
\begin{aligned}
\Psi_{\mathbf n,\mathrm{ref}}^{(0)}(\{z_i\})
&=
{\cal S}\!\left[
\prod_{i=1}^{N}\varphi_{m_i}^{(N_d)}(z_i)
\right]
F_{\rm CM}(Z)
\prod_{i<j}J_2(z_i-z_j),\\
\Psi_{\mathbf n}^{(0)}(\{x_i,y_i\})
&=
\left[
\Psi_{\mathbf n,\mathrm{ref}}^{(0)}(\{z_i\})
\right]^* .
\end{aligned}
\label{eq:CB_ansatz_main}
\end{equation}
Our positive-flux convention follows the lattice Peierls convention of
Eq.~\eqref{eq:peierls_phase}, whereas the first line of
Eq.~\eqref{eq:CB_ansatz_main} constructs a reference trial seed from
conventional theta-function building blocks of the opposite magnetic
chirality~\cite{HaldaneRezayi1985,GreiterSchnellsThomale2016}. These blocks
explicitly implement the torus magnetic boundary conditions and vortex
structure and the numerical seed is their complex conjugate, as shown in the
second line. Complex conjugation reverses chirality and merely relabels the
magnetic-translation sectors, leaving energies, state counting, and
trial-space rank unchanged.

In Eq.~\eqref{eq:CB_ansatz_main} \(\varphi_m^{(N_d)}\) are analytic single-particle torus orbitals at
reduced flux, rather than numerical Hofstadter eigenstates
(cf.~\cite{SupplementalMaterial}). Together with the Jastrow factor, they make
the reduced-flux composite-boson structure and magnetic boundary conditions
explicit, while providing analytic amplitudes suitable for VMC.  The
two-vortex Jastrow factor is
\begin{equation}
J_2(z_i-z_j)=
\vartheta_{\frac12,\frac12}^{\,2}
\!\left(
\frac{z_i-z_j}{L_1}\middle|\tau
\right).
\label{eq:Jastrow_main}
\end{equation}
It is the square of the odd Jacobi theta function, the torus analogue of
\((z_i-z_j)^2\), and attaches two vortices to each
boson~\cite{Jain2007,HaldaneRezayi1985,GreiterSchnellsThomale2016}.  The Gaussian
factors restoring the physical magnetic length are given in the
Supplemental Material~\cite{SupplementalMaterial}.

In the reference analytic convention, the center-of-mass factor obeys
\begin{equation}
\begin{aligned}
F_{\rm CM}(Z+L_1)&=F_{\rm CM}(Z),\\
F_{\rm CM}(Z+\ii L_2)&=
e^{-\ii2\pi(2Z/L_1+\tau)}F_{\rm CM}(Z).
\end{aligned}
\label{eq:FCM_BC_main}
\end{equation}
A convenient solution is
\(F_{\rm CM}(Z)=\vartheta_{a,0}(2Z/L_1|2\tau)\), with an appropriate
characteristic \(a\) (in our numerics we choose \(a=0\)).  The factor of two follows from the boundary
conditions of \(J_2\); it is not itself the torus center-of-mass degeneracy.

For ansatz-subspace diagonalization and VMC, each physical seed generates the
trial family through repeated application of \(T_{\rm CM}\), the collective
magnetic translation shifting all particles by one lattice spacing along
\(y\) in the physical \(+\phi\) convention:
\begin{equation}
{\cal A}_{\mathbf n}
=
\left\{(T_{\rm CM})^q\Psi_{\mathbf n}^{(0)}\,\big|\,
q=0,\ldots,q_{\rm COM}-1\right\}.
\label{eq:TCM_family_main}
\end{equation}
Each \({\cal A}_{\mathbf n}\) contains \(q_{\rm COM}\) translated states,
where \(q_{\rm COM}\) is specified below; all \(C_{\rm B}\) occupation
patterns thus produce \(C_{\rm B}q_{\rm COM}\) raw states. Organizing the
reduced-flux occupation patterns into cyclic orbits predicts that this raw
trial collection is overcomplete. Once the full center-of-mass multiplet is included, this overcompleteness reflects linear dependencies among trial families associated with occupation patterns related by a common shift of all reduced-flux orbital labels. Writing \(g=\gcd(N,N_\phi)=\gcd(N,N_d)\), the cyclic-orbit structure predicts \(D_{\rm int}=C_{\rm B}g/N_d\) independent internal seeds, while \(q_{\rm COM}=N_\phi/g\). Thus
\(D_{\rm int}q_{\rm COM}=C_{\rm B}N_\phi/N_d\), reproducing
Eq.~\eqref{eq:CB_counting_formula}. The number of nonzero eigenvalues of the trial-state overlap matrix confirms this physical rank; an orbit-resolved analysis further verifies the predicted reduction within each cyclic occupation orbit~\cite{SupplementalMaterial}. The center-of-mass structure is also reflected in the variational spectra: diagonalizing the lowest-band-projected HBH Hamiltonian within the CB space yields energy subgroups whose multiplicities are integer multiples of \(q_{\rm COM}\).

{\it Fractional density depletion, quasihole braiding, and the many-body Chern number---}
We characterize the unpinned low-energy manifold by its many-body Chern
number and probe additional flux excitations through density depletion and
braiding. In the pinned calculations, \(N_{\rm loc}\) added flux quanta are
localized by repulsive potentials. We refer to the resulting localized
quasihole-like excitations simply as quasiholes below. Applying Eq.~\eqref{eq:CB_counting_formula} with
\(N_\phi\to N_\phi'\equiv N_\phi-N_{\rm loc}\) and
\(N_d\to N_d-N_{\rm loc}\) gives the pinned-manifold dimension \(D\)~\cite{ZeybekUmucalilar2022},
to be distinguished from the unpinned dimension \(\mathcal D\), and we
define \(\nu_{\rm eff}=N/N_\phi'\).

For the charge diagnostic, let
\(\Psi_{\rm pin}=(|\psi_1\rangle,\ldots,|\psi_D\rangle)\)
denote the matrix whose columns form an orthonormal basis for the pinned low-energy manifold and let
\(\widetilde n_i\) denote the site-occupation operator projected onto the
lowest band.  We calculate
\begin{equation}
\begin{aligned}
\bar n_i&=\frac{1}{D}\operatorname{Tr}
\!\left(\Psi_{\rm pin}^{\dagger}
\widetilde n_i\Psi_{\rm pin}\right),\\
Q(R)&=\sum_{i\in\mathcal R(R)}(n_0-\bar n_i),
\qquad n_0=\nu_{\rm eff}\phi ,
\end{aligned}
\label{eq:fractional_depletion}
\end{equation}
where \(\mathcal R(R)\) is the disk of radius \(R\) centered on the pin.
Once it encloses the quasihole core, the plateau of \(Q(R)\) gives the effective localized quasihole charge~\cite{Raciunas2018,Macaluso2020,WangDongEckardt2022}.

For braiding, we follow the \(D\)-dimensional pinned subspace by discrete
parallel transport. At each step, the singular-value decomposition of the
overlap matrix aligns the instantaneous orthonormal basis with the preceding
one~\cite{Vanderbilt2018,SupplementalMaterial}. Let \(\Psi_0\) contain the
initial basis and \(\Psi_L\) its parallel-transported basis after the closed
path. Their overlap defines the \(D\times D\) non-Abelian Berry matrix
\(\mathcal B\)~\cite{Berry1984,WilczekZee1984}. Denoting its eigenvalues by
\(\lambda_a\), we write
\begin{equation}
\begin{aligned}
\mathcal B&=\Psi_L^\dagger\Psi_0,\\
\theta_a&=\frac{1}{\pi}\arg\lambda_a,\qquad
\overline{\theta}_{\rm B}=\frac{1}{D}\sum_{a=1}^{D}\theta_a .
\end{aligned}
\label{eq:berry_phase_average}
\end{equation}
Here \(\theta_a\) and \(\overline{\theta}_{\rm B}\) are quoted in units of
\(\pi\) modulo \(2\); the mean uses
eigenphases on a consistent principal or adjacent branch. We use the
cross-shaped commutator loop
\(\mathcal C=T_2^{-1}T_1^{-1}T_2T_1\), with one quasihole fixed during each
transport segment~\cite{WenDagottoFradkin1990}.  Its forward and reverse segments cancel the Aharonov--Bohm phase accumulated
from motion in the external magnetic field, isolating the mutual braiding
contribution. We compare the resulting mean phase with
\(\overline{\theta}_{\rm B}=2\nu_{\rm eff}\). For a standard Laughlin
quasihole at \(\nu=1/m\), the full-braid phase is \(2/m\) in the same
units~\cite{Arovas1984}, coinciding with this value when
\(\nu_{\rm eff}=1/m\).

For the unpinned system, we compute the many-body Chern number
\(C_{\rm MB}\) of the full \(\mathcal D\)-state manifold from determinant
link variables on a mesh of boundary
twists~\cite{NiuThoulessWu1985,FukuiHatsugaiSuzuki2005,
KudoKariyadoHatsugai2017,SupplementalMaterial}.  Unlike the charge and braiding calculations, this
calculation uses no pinning potential and tests
\(C_{\rm MB}/\mathcal D=\nu=N/N_\phi\).

{\it Numerical Results---}
Our main calculations project to the lowest band of the single-particle
Hofstadter spectrum. The site-basis trial states are transformed to the projected Fock basis
using bosonic permanents~\cite{SupplementalMaterial}. We set \(t=1\) and quote energies in units of \(t\).
Projected calculations use \(U/t=2\), whereas real-space VMC uses the
hard-core limit \(U/t\to\infty\). We generally take \(\phi=1/L_1\) and set
\(L_1=L_2\) for unpinned systems. To create localized quasiholes at the same
effective filling, we keep \(L_1\) and \(\phi\) fixed and set \(L_2=L_1+1\)
for density depletion and \(L_2=L_1+2\) for braiding. The added fluxes are
pinned, giving \(N_\phi'\equiv N_\phi-N_{\rm loc}=L_1\) and
\(\nu_{\rm eff}=N/N_\phi'\).

\begin{figure}[t]
\centering
\includegraphics[width=\columnwidth]{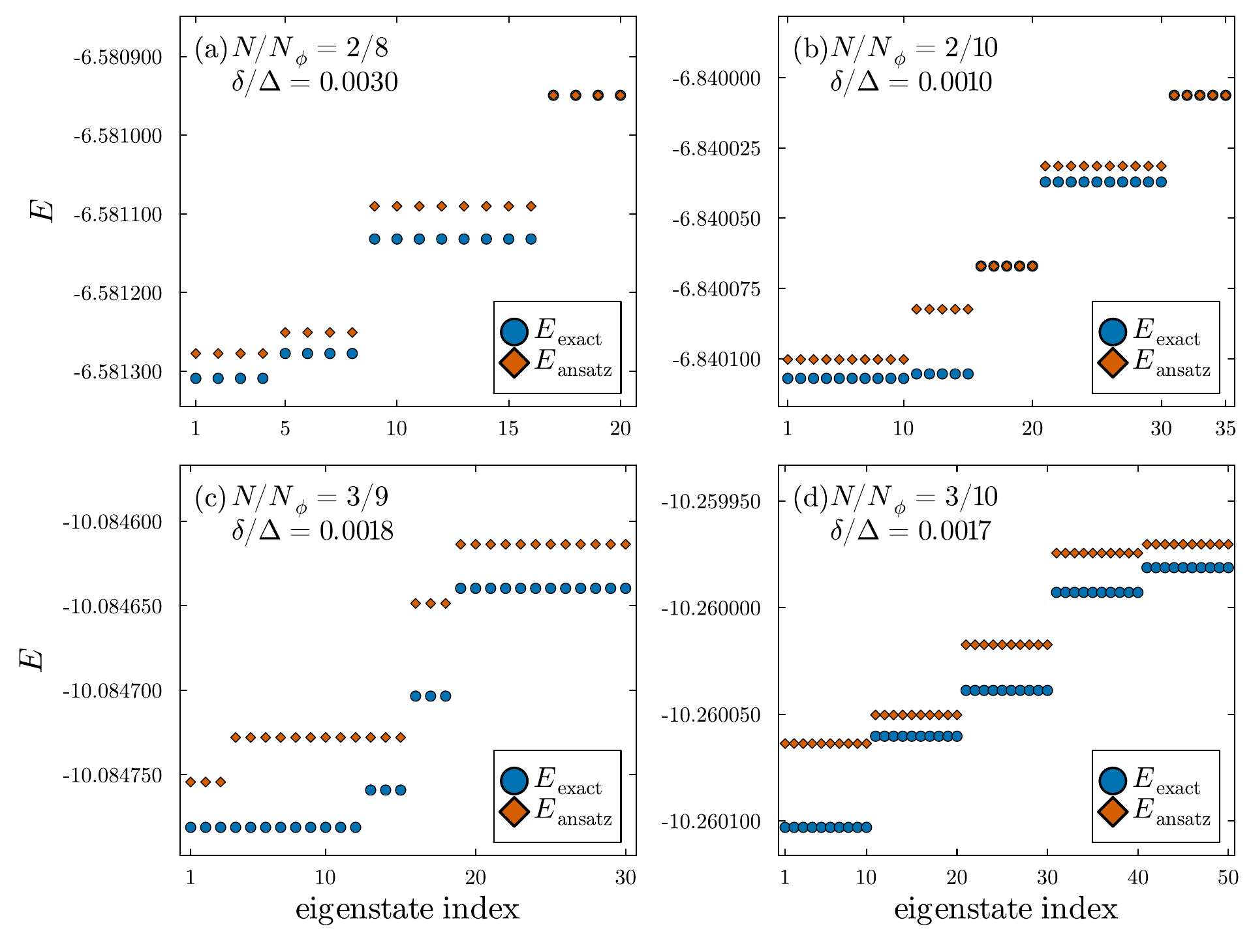}
\caption{\label{fig:energy_spectra}Exact lowest-band-projected
\((E_{\rm exact})\) and ansatz-subspace \((E_{\rm ansatz})\) spectra for
\((N,N_\phi)=(2,8),(2,10),(3,9),(3,10)\), corresponding to
\(\nu=1/4,1/5,1/3,3/10\).  The respective raw-state counts and physical
ranks \((N_{\rm raw},\mathcal D)\) are
\((40,20),(105,35),(30,30),(200,50)\); the ranks equal both the observed
quasidegeneracies and Eq.~\eqref{eq:CB_counting_formula}.
}
\end{figure}

Fig.~\ref{fig:energy_spectra} compares the exact and CB-trial-space spectra of
the lowest-band-projected Hamiltonian. The energy-subgroup multiplicities
agree in every panel, partly reflecting the magnetic-translation symmetry
shared by the two calculations, and several energies coincide in (a) and (b);
only in (c) is the ordering of the two lowest subgroups reversed. Because
\(\delta/\Delta\leq3.0\times10^{-3}\) in all panels, with \(\delta\) the
manifold bandwidth and \(\Delta\) the gap above it, this reversal concerns
minute splittings within a single well-isolated manifold. The Supplemental Material~\cite{SupplementalMaterial} presents
orbit-resolved counting, numerical-rank and subspace-fidelity analyses,
and the observed energy-subgroup multiplicities.  Direct CB--exact
subspace fidelities larger than \(0.9996\), together with the close
spectra, show that the trial span accurately describes the isolated
low-energy manifold.

For the larger unpinned systems \((N,N_\phi)=(5,11)\) and \((6,14)\),
VMC evaluation of the ansatz matrices recovers the predicted ranks
\(\mathcal D=11\) and \(49\). The hard-core real-space VMC spectra remain
narrow, with spreads \(0.052\Delta\) and \(0.113\Delta\) for the two
systems, respectively, and all ordered levels lie within \(0.122\Delta\)
of their \(U/t=2\) lowest-band-projected counterparts. Thus, both the
predicted rank and the narrow-manifold energetics persist without explicit
lowest-band projection. Leave-one-bin-out jackknife
errors~\cite{SupplementalMaterial,Efron1982Jackknife} are below
\(0.006\Delta\), substantially smaller than the reported spectral scales.

Beyond their counting and energetics, we probe the FQH character of these
manifolds through complementary calculations of their many-body Chern numbers,
fractional density depletion, and quasihole-like braiding. For all unpinned case studied (cf.~\cite{SupplementalMaterial}), spanning \(N=2\)--5,
\(N_\phi=5\)--12, and \(\mathcal D=5\)--105, we obtain the expected many-body Chern
number,
\[
C_{\rm MB}=\binom{N_d+N-1}{N_d},
\]
and hence \(C_{\rm MB}/\mathcal D=\nu\).

Full-manifold averaging is essential: the charge and braiding diagnostics use
all \(D\) pinned states, while \(C_{\rm MB}/\mathcal D\) involves the complete
\(\mathcal D\)-state unpinned manifold. Although large quasihole manifolds are routinely characterized through energy
and entanglement spectra~\cite{RegnaultBernevig2011,BernevigRegnault2012},
direct charge and braiding calculations have generally focused on isolated
states or small pinned multiplets~\cite{LiuBhattRegnault2015,
JaworowskiRegnaultLiu2019,Raciunas2018,Macaluso2020,
WangDongEckardt2022,ZeybekUmucalilar2022}; here the corresponding
pinned-manifold dimensions reach \(D=105\) and \(D=11\), respectively.

\begin{figure}[t]
\centering
\includegraphics[width=\columnwidth]{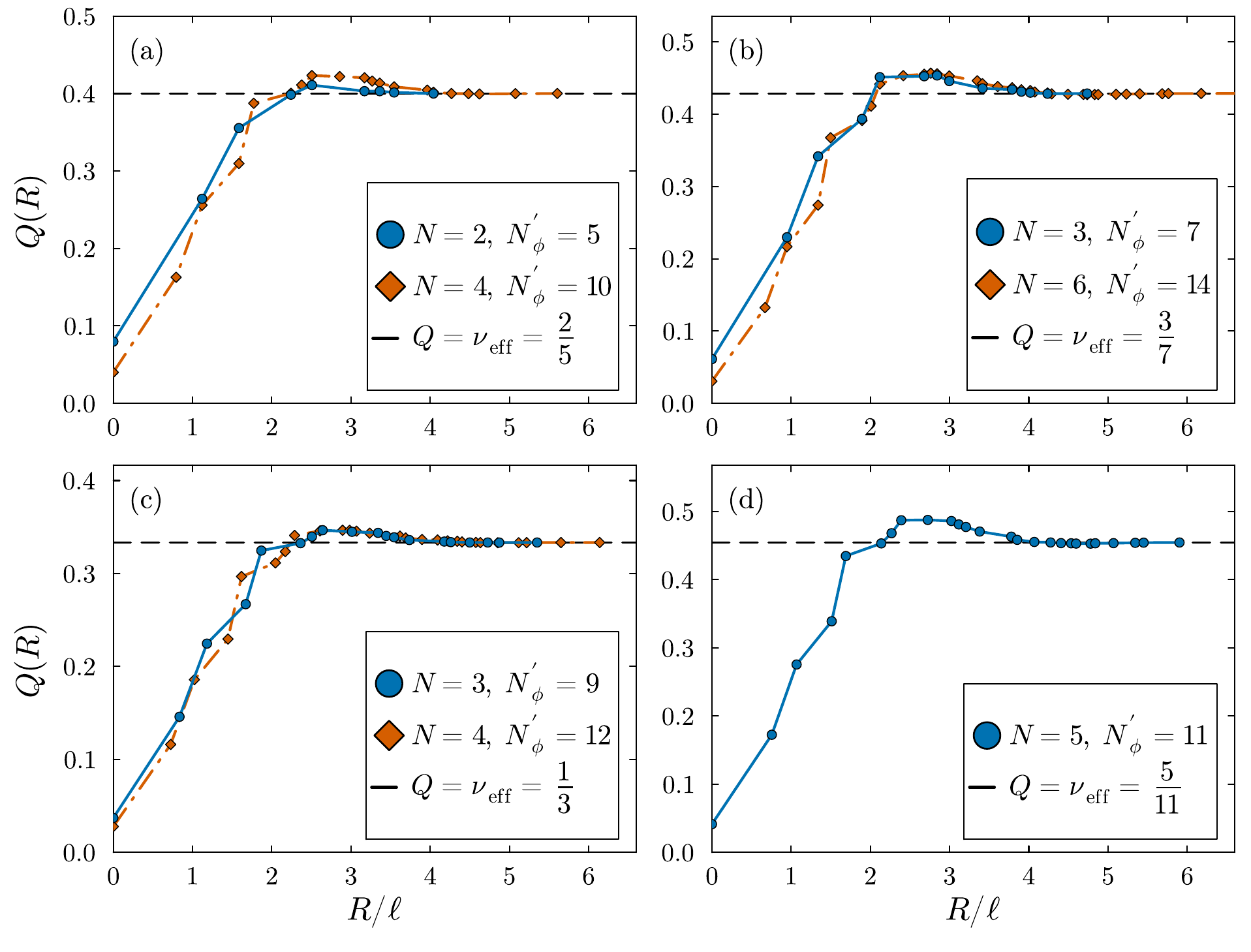}
\caption{\label{fig:charge_depletion}Accumulated density depletion versus
\(R/\ell\) around one pinned quasihole.  Panels (a)--(d) show
\((N,N_\phi')=(2,5),(4,10)\);
\((3,7),(6,14)\); \((3,9),(4,12)\); and \((5,11)\), with
\(\nu_{\rm eff}=2/5,3/7,1/3,\) and \(5/11\), respectively.  Dashed lines
mark \(Q=\nu_{\rm eff}\); the filling in (d) is braided in
Fig.~\ref{fig:braiding}. All panels use a pinning strength \(V/t=1\).}
\end{figure}

Fig.~\ref{fig:charge_depletion} shows the accumulated depletion around one
pin.  Expressing \(R\) in magnetic-length units makes systems at the same
\(\nu_{\rm eff}\), but with different \(N\) and \(\phi\), approach the
same plateau: once the quasihole core is enclosed,
\(Q(R)=\nu_{\rm eff}\) within numerical accuracy.

\begin{figure}[t]
\centering
\includegraphics[width=\columnwidth]{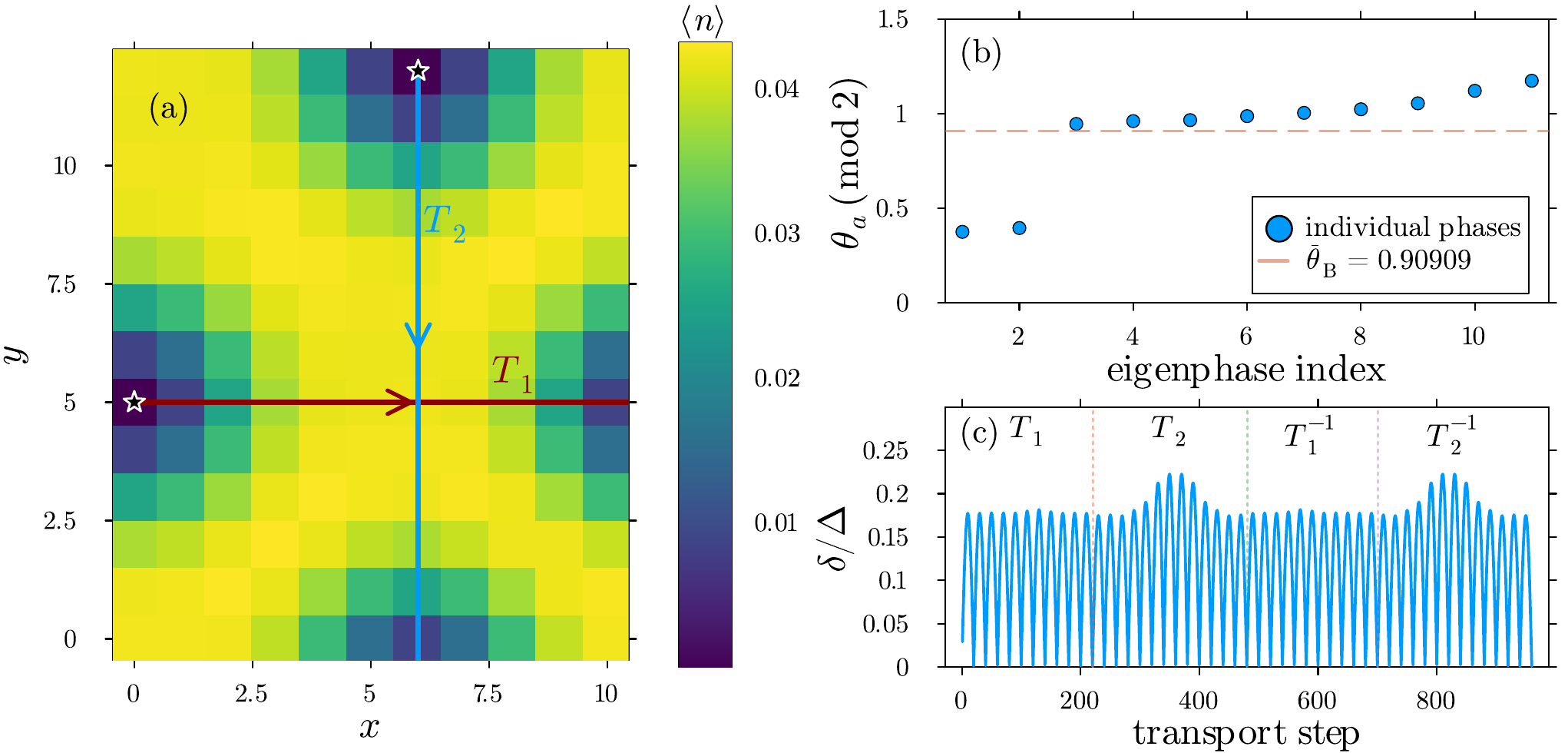}
\caption{\label{fig:braiding}AB-free quasihole braid at
\(\nu_{\rm eff}=5/11\) with identical pin strengths \(V_1=V_2=0.8t\). (a) Initial subspace-averaged density; stars and
arrows mark the quasihole positions and cross-shaped paths.  (b) Berry
eigenphases and their mean, in units of \(\pi\); the dashed line is
\(2\nu_{\rm eff}=10/11\).  (c) \(\delta/\Delta\) along
\(T_1,T_2,T_1^{-1},T_2^{-1}\); dotted lines separate the four branches. The minima of \(\delta/\Delta\) coincide with path points at which the quasiholes are centered on lattice sites. Each elementary displacement of a quasihole is implemented by transferring the pinning potential between neighboring sites in discrete steps, lowering its strength at the departure site while raising it at the destination site.}
\end{figure}

For the braiding of two quasiholes in an \(N=5\), \(11\times13\) system
(\(\phi=1/11\), \(D=11\)), Fig.~\ref{fig:braiding} gives
\(\overline{\theta}_{\rm B}=10/11=2\nu_{\rm eff}\), while the
manifold remains isolated throughout the path
(\(\delta/\Delta<0.23\)). Owing to the commutator structure of the
AB-free loop, the numerically evaluated Berry matrix satisfies
\(\det\mathcal B\approx1\), restricting the mean of either the principal
phases or any consistently chosen adjacent branches to
\(\overline{\theta}_{\rm B}\approx2m/D\), \(m\in\mathbb Z\).
This constraint does not determine \(m\); the nontrivial result of the
present calculation is its selection of \(m=5\). In the Supplemental
Material~\cite{SupplementalMaterial}, we show for a smaller system that
the corresponding physical integer remains correctly selected upon
introducing a random background potential and asymmetric pinning. This robustness supports the interpretation of the selected phase as a
quasihole-like braiding response of the localized excitations rather than an
artifact of residual spatial symmetries in the clean protocol.

{\it Conclusion---}
Our results provide a microscopic composite-boson wave-function description
of the quasidegenerate, lattice-split Laughlin-quasihole manifolds of the
low-flux HBH model at generic fillings \(\nu<1/2\), going beyond
their previous characterization through exclusion-rule counting. Organizing
the reduced-flux occupation patterns into cyclic orbits and including the
center-of-mass multiplicity yields a physical-rank prediction that is confirmed
by the overlap matrices and matches the observed manifold dimension.
Separately, diagonalization within the resulting ansatz span reproduces the
observed energy-subgroup multiplicities and closely follows the exact projected
spectra, while VMC extends the rank and energetic analysis to larger systems.

Complementary diagnostics reveal a coherent fractional topological response.
The unpinned manifolds satisfy \(C_{\rm MB}/\mathcal D=\nu\), while an
additional localized flux excitation carries an effective quasihole charge
\(Q\simeq\nu_{\rm eff}\). An AB-free braid gives a phase consistent with
\(2\nu_{\rm eff}\), with disorder and asymmetric-pinning checks in a smaller
system supporting its quasihole-like interpretation.

Natural next steps are to test whether the same reduced-flux organization
persists in flat-band lattice models such as the Kapit--Mueller model and in
continuum bosons with short-range interactions, and to determine the origin of
the residual finite-size energy-subgroup splittings. Longer-range interactions
may also provide a natural route to selecting gapped finite-degeneracy sectors
from these manifolds at sub-half filling, as suggested previously for the
bosonic \(\nu=1/4\) state~\cite{Hafezi2007}. These directions would clarify how
far the present composite-boson organization extends beyond the low-flux HBH
setting.

{\it Acknowledgments---}
R.O.U. acknowledges support through the 2023 T\"UB\.{I}TAK Incentive Award. The authors acknowledge the use of OpenAI's ChatGPT-5-series models, directed through task-specific prompts, for assistance with manuscript refinement and with drafting and debugging portions of the numerical code, including VMC routines. The resulting text and code were critically reviewed and verified by the authors, who take full responsibility for the calculations, interpretations, and conclusions.

\bibliographystyle{apsrev4-2}
\bibliography{CB_ansatz_refs}
\end{document}


\title{Supplemental Material}
\maketitle
\section{Magnetic-translation conventions and boundary conditions of the composite-boson ansatz}
\label{sec:magnetic_translation_CB_BCs}

This section distinguishes the magnetic-translation convention of the physical Hofstadter problem from the opposite-chirality reference convention used to construct the torus wave functions~\cite{HaldaneRezayi1985,GreiterSchnellsThomale2016}.  We then derive the boundary conditions imposed on the center-of-mass factor in the reference composite-boson ansatz and explain how the resulting states are mapped to the physical convention.

We use a rectangular torus with periods $L_1$ and $\ii L_2$, complex coordinate $z=x+\ii y$, modular parameter
\begin{equation}
    \tau=\ii\frac{L_2}{L_1}.
\end{equation}
The physical Hofstadter Hamiltonian in the main text has positive flux per
plaquette $\phi$ and uses the Landau gauge
\begin{equation}
    \vv A=-By\,\hat{\vv x}.
\end{equation}
With lengths measured in units of the lattice constant,
\begin{equation}
    \phi=\frac{\Nphi}{L_1L_2}>0,
    \qquad
    \frac{1}{\ell^2}=2\pi\phi
    =\frac{2\pi\Nphi}{L_1L_2}.
\end{equation}
Here $\ell$ denotes the magnetic length. The theta-function construction below instead defines a reference
state of the opposite magnetic chirality, equivalently at flux $-\phi$.
We keep $\Nphi$, $\phi$, and $\ell$ positive as magnitudes; the reference
chirality is specified by the magnetic boundary conditions.

\subsection{Reference magnetic boundary conditions}
\label{subsec:reference_translation_convention}

For the opposite-chirality reference state, we define the two
full-period magnetic translations directly by
\begin{align}
    \bigl(t_{1,{\rm ref}}\psi\bigr)(x,y)
    &=\psi(x+L_1,y),\\
    \bigl(t_{2,{\rm ref}}\psi\bigr)(x,y)
    &=\ee^{+\ii2\pi\Nphi x/L_1}\psi(x,y+L_2).
    \label{eq:reference_period_translations}
\end{align}
Magnetic translations by general displacements need not commute.  For
translations by the two full torus periods, however,
\begin{equation}
    t_{1,{\rm ref}}t_{2,{\rm ref}}
    =\ee^{\ii2\pi\Nphi}t_{2,{\rm ref}}t_{1,{\rm ref}}
    =t_{2,{\rm ref}}t_{1,{\rm ref}},
\end{equation}
where \(L_1L_2=2\pi\Nphi\ell^2\).  Since the torus contains an integer
number \(\Nphi\) of flux quanta, the phase factor is unity and the
full-period magnetic translations commute.

\subsection{Single-particle boundary conditions}
\label{subsec:single_particle_BC}

For completeness, we rederive the standard Haldane--Rezayi
single-particle boundary conditions in the reference gauge and chirality
convention used here~\cite{HaldaneRezayi1985}.  Besides fixing our
conventions, these conditions determine the quasiperiodicity of the
reduced-flux single-particle orbitals entering the composite-boson
ansatz below.

Let
\begin{equation}
    \psi(z)=f(z)\ee^{-y^2/(2\ell^2)}
    =f(z)\ee^{-\pi\phi y^2}.
\end{equation}
Here $f(z)$, the non-Gaussian factor, is holomorphic (complex-analytic in
$z$); the full state $\psi(z)$ is not holomorphic because it also contains
the Gaussian.
The reference periodic magnetic boundary conditions are
\begin{equation}
    t_{1,{\rm ref}}\psi=\ee^{\ii\phix}\psi,
    \qquad
    t_{2,{\rm ref}}\psi=\ee^{\ii\phiy}\psi .
\end{equation}
The $L_1$ translation directly gives
\begin{equation}
    \frac{f(z+L_1)}{f(z)}=\ee^{\ii\phix}.
    \label{eq:single_f_L1}
\end{equation}
For the $L_2$ translation, using Eq.~\eqref{eq:reference_period_translations},
\begin{align*}
    t_{2,{\rm ref}}\psi(z)
    &=f(z+\ii L_2)
      \ee^{-\pi\phi(y+L_2)^2}
      \ee^{+\ii2\pi\phi xL_2}\\
    &=f(z+\ii L_2)
      \ee^{-\pi\phi y^2}
      \ee^{\ii\pi\Nphi(2z/L_1+\tau)} .
\end{align*}
Hence
\begin{equation}
    \frac{f(z+\ii L_2)}{f(z)}
    \ee^{\ii\pi\Nphi(2z/L_1+\tau)}
    =\ee^{\ii\phiy}.
    \label{eq:single_f_L2}
\end{equation}
These are the standard Haldane--Rezayi quasiperiodicity conditions for
the holomorphic factor on a rectangular torus.  For a system pierced by
$\Nphi$ flux quanta, analyticity further implies that $f(z)$ has
$\Nphi$ zeros in a fundamental region and that the space of solutions
has dimension $\Nphi$~\cite{HaldaneRezayi1985}.  The corresponding
functions can be represented conveniently in terms of theta functions.

We use the convention
\begin{equation}
    \vartheta_{a,b}(u|\tau)
    =\sum_{n=-\infty}^{\infty}
    \ee^{\ii\pi\tau(n+a)^2+2\pi\ii(n+a)(u+b)},
\end{equation}
with the quasiperiodicities
\begin{equation}
\begin{aligned}
    \vartheta_{a,b}(u+1|\tau)
    &=\ee^{2\pi\ii a}\vartheta_{a,b}(u|\tau),
    \\
    \vartheta_{a,b}(u+\tau|\tau)
    &=\ee^{-\ii\pi\tau-2\pi\ii(u+b)}
      \vartheta_{a,b}(u|\tau).
\end{aligned}
\label{eq:theta_quasiperiodicity}
\end{equation}
For zero twists, a convenient basis of the $\Nphi$ independent
single-particle solutions is
\begin{equation}
    f_m^{(\Nphi)}(z)
    =
    \vartheta_{\frac{m}{\Nphi},0}
    \!\left(
        \frac{\Nphi z}{L_1}
        \middle|
        \Nphi\tau
    \right),
    \qquad
    m=0,1,\ldots,\Nphi-1 .
    \label{eq:single_particle_theta_basis}
\end{equation}
Indeed, Eq.~\eqref{eq:theta_quasiperiodicity} immediately gives
Eqs.~\eqref{eq:single_f_L1} and \eqref{eq:single_f_L2} with
$\phix=\phiy=0$.  The corresponding full single-particle orbitals are
\begin{equation}
    \varphi_{m,\mathrm{ref}}^{(\Nphi)}(z)
    =
    f_m^{(\Nphi)}(z)\,
    \ee^{-\pi\phi y^2}.
    \label{eq:single_particle_torus_orbitals}
\end{equation}
The reduced-flux orbitals used below have the same form with
$\Nphi$ replaced by $\Nd$ and with the corresponding reduced flux
density
\begin{equation}
    \phi_d=\frac{\Nd}{L_1L_2}.
\end{equation}

For either \(M=\Nphi\) or \(M=\Nd\), the orbital index \(m\) has a
direct magnetic-translation meaning.  In the present Landau gauge, a
translation along direction \(x\) carries no additional gauge phase,
\[
    \bigl(t_{1,{\rm ref}}(\delta)\psi\bigr)(x,y)
    =\psi(x+\delta,y).
\]
Equation~\eqref{eq:theta_quasiperiodicity} therefore gives
\begin{equation}
    t_{1,{\rm ref}}\!\left(\frac{L_1}{M}\right)
    \varphi_{m,{\rm ref}}^{(M)}
    =
    \ee^{2\pi\ii m/M}\varphi_{m,{\rm ref}}^{(M)} .
    \label{eq:reference_magnetic_momentum}
\end{equation}
Thus \(m\pmod M\) is the reference magnetic-momentum label, defined by
this translation eigenphase.

\subsection{Laughlin/Jastrow factor and the center-of-mass boundary conditions}
\label{subsec:Laughlin_BC}

We next apply the same theta-function quasiperiodicities to the
many-particle Laughlin/Jastrow factor.  The standard reference torus
Laughlin-type form at filling $\nu=1/m$ is~\cite{HaldaneRezayi1985}
\begin{equation}
    \Psi_{m,{\rm ref}}(\{z_i\})=
    \FCM(Z)
    \prod_{i<j}
    \vtheta^m\!\left(\frac{z_i-z_j}{L_1}\middle|\tau\right)
    \ee^{-\sum_i y_i^2/(2\ell^2)},
    \qquad Z=\sum_i z_i,
    \label{eq:Laughlin_torus_form}
\end{equation}
where
$\vtheta(u|\tau)$
is the odd Jacobi theta function.  From
Eq.~\eqref{eq:theta_quasiperiodicity},
\begin{equation}
    \vtheta(u+1|\tau)=-\vtheta(u|\tau).
\end{equation}
Translating one particle, say $z_1$, by $L_1$ therefore gives
$N-1$ theta-function phases, each raised to the $m$th power:
\begin{equation}
    \prod_{j=2}^N
    \vtheta^m\!\left(
        \frac{z_1+L_1-z_j}{L_1}\middle|\tau
    \right)
    =
    (-1)^{m(N-1)}
    \prod_{j=2}^N
    \vtheta^m\!\left(
        \frac{z_1-z_j}{L_1}\middle|\tau
    \right).
\end{equation}
Using $\Nphi=mN$, this phase may be written as
$(-1)^{\Nphi-m}$, and the periodic boundary condition gives
\begin{equation}
    \frac{\FCM(Z+L_1)}{\FCM(Z)}
    =(-1)^{\Nphi-m}\ee^{\ii\phix}.
    \label{eq:FCM_Laughlin_L1}
\end{equation}

For the $L_2$ translation, Eq.~\eqref{eq:theta_quasiperiodicity} gives, for the $N-1$ factors containing $z_1$,
\begin{widetext}
\begin{align*}
&\prod_{j=2}^N
\vtheta^m\!\left(\frac{z_1+\ii L_2-z_j}{L_1}\middle|\tau\right)\\
&\quad=
\ee^{-\ii\pi\tau m(N-1)}
\ee^{-\ii\pi m(N-1)}
\ee^{-\frac{2\pi\ii m}{L_1}[(z_1-z_2)+\cdots+(z_1-z_N)]}
\prod_{j=2}^N
\vtheta^m\!\left(\frac{z_1-z_j}{L_1}\middle|\tau\right)\\
&\quad=
\ee^{-\ii\pi\Nphi(2z_1/L_1+\tau)}
(-1)^{\Nphi-m}
\ee^{\ii\pi m(2Z/L_1+\tau)}
\prod_{j=2}^N
\vtheta^m\!\left(\frac{z_1-z_j}{L_1}\middle|\tau\right).
\end{align*}
\end{widetext}
The Gaussian and magnetic-translation phase give the compensating factor
\begin{equation}
    \ee^{-\pi\phi(y_1+L_2)^2}
    \ee^{+\ii2\pi\phi x_1L_2}
    =\ee^{-\pi\phi y_1^2}
    \ee^{\ii\pi\Nphi(2z_1/L_1+\tau)}.
\end{equation}
The $z_1$-dependent factors cancel, and the remaining condition on the center-of-mass part is
\begin{equation}
    \frac{\FCM(Z+\ii L_2)}{\FCM(Z)}
    \ee^{\ii\pi m(2Z/L_1+\tau)}
    =(-1)^{\Nphi-m}\ee^{\ii\phiy}.
    \label{eq:FCM_Laughlin_L2}
\end{equation}
Equations~\eqref{eq:FCM_Laughlin_L1} and \eqref{eq:FCM_Laughlin_L2} are the Haldane--Rezayi center-of-mass boundary conditions for the reference Laughlin/Jastrow form.

\subsection{Composite-boson ansatz}
\label{subsec:CB_ansatz_BC}

The reference composite-boson variational ansatz used in the main text is
\begin{equation}
    \Psi_{\mathrm{CB,ref}}(\{z_i\}) =
    \mathcal S\!\left[\prod_{i=1}^N\varphi_{m_i}(z_i)\right]
    \FCM(Z)
    \left[
    \prod_{i<j}
    \vtheta\!\left(\frac{z_i-z_j}{L_1}\middle|\tau\right)
    \ee^{-\sum_i y_i^2/(2\ell'^2)}
    \right]^2 .
    \label{eq:CB_ansatz}
\end{equation}
Here the symmetrization acts only on the single-particle factor.  The remaining factors are already symmetric under particle exchange: the squared theta-function product is symmetric and the Gaussian and center-of-mass factors are symmetric.  Consequently, it is sufficient to derive the boundary conditions for one unsymmetrized term in the symmetrized single-particle product.

The single-particle reference orbitals $\varphi_{m_i}$ are the
theta-function orbitals of Eq.~\eqref{eq:single_particle_torus_orbitals},
now taken at the reduced flux
\begin{equation}
    \Nd=\Nphi-2N.
    \label{eq:reduced_flux}
\end{equation}
The orbitals already contain the Gaussian factor corresponding to $\Nd$ fluxes, while the full ansatz should have the Gaussian corresponding to the original magnetic length $\ell$.  Therefore $\ell'$ is fixed by
\begin{equation}
    \frac{1}{2\ell_d^{2}}+\frac{1}{\ell'^2}
    =\frac{1}{2\ell^2}.
    \label{eq:ellprime_condition}
\end{equation}
Using $1/\ell^2=2\pi\Nphi/(L_1L_2)$ and $1/\ell_d^{2}=2\pi\Nd/(L_1L_2)$ gives
\begin{equation}
    \frac{1}{\ell'^2}=\frac{2N\pi}{L_1L_2}.
    \label{eq:ellprime_value}
\end{equation}
Thus the explicit Gaussian produced by the squared bracket in Eq.~\eqref{eq:CB_ansatz} is
\begin{equation}
    \ee^{-\frac{2N\pi}{L_1L_2}\sum_i y_i^2}.
    \label{eq:explicit_CB_gaussian}
\end{equation}

We set the many-body twist phases to zero in this subsection.  Nonzero twists can be restored by multiplying the right-hand sides of the corresponding boundary conditions by $\ee^{\ii\phix}$ and $\ee^{\ii\phiy}$.

\subsubsection{Translation by $L_1$}

Under $z_1\mapsto z_1+L_1$, every theta function involving $z_1$ acquires a sign, but the square removes it:
\begin{equation}
    \vtheta^2\!\left(\frac{z_1+L_1-z_j}{L_1}\middle|\tau\right)
    =\vtheta^2\!\left(\frac{z_1-z_j}{L_1}\middle|\tau\right).
\end{equation}
Hence the $L_1$ boundary condition of the full ansatz reduces to
\begin{equation}
    \varphi_{m_1}(z_1+L_1)\FCM(Z+L_1)
    =\varphi_{m_1}(z_1)\FCM(Z).
    \label{eq:CB_intermediate_L1}
\end{equation}
The reduced-flux single-particle orbitals are chosen to be periodic along $L_1$,
\begin{equation}
    \varphi_{m_1}(z_1+L_1)=\varphi_{m_1}(z_1),
\end{equation}
so the center-of-mass factor must obey
\begin{equation}
    \boxed{\FCM(Z+L_1)=\FCM(Z).}
    \label{eq:CB_FCM_L1}
\end{equation}

\subsubsection{Translation by $\ii L_2$}

The theta-function part gives the $m=2$ specialization of the calculation in Sec.~\ref{subsec:Laughlin_BC}:
\begin{align}
&\prod_{j=2}^N
\vtheta^2\!\left(\frac{z_1+\ii L_2-z_j}{L_1}\middle|\tau\right)
\notag\\
&\quad=
\ee^{-\ii2\pi N(2z_1/L_1+\tau)}
\ee^{\ii2\pi(2Z/L_1+\tau)}
\prod_{j=2}^N
\vtheta^2\!\left(\frac{z_1-z_j}{L_1}\middle|\tau\right).
\label{eq:CB_theta_L2_factor}
\end{align}
The explicit Gaussian in Eq.~\eqref{eq:explicit_CB_gaussian}, together with the reference magnetic-translation phase, gives
\begin{align}
&\ee^{-\frac{2N\pi}{L_1L_2}(y_1+L_2)^2}
\ee^{+\ii2\pi\phi x_1L_2}
\notag\\
&\quad=
\ee^{-\frac{2N\pi}{L_1L_2}y_1^2}
\ee^{\ii2\pi N(2z_1/L_1+\tau)}
\ee^{\ii\pi\Nd(2x_1/L_1)}.
\label{eq:CB_gaussian_L2_factor}
\end{align}
The first exponential on the right-hand side of Eq.~\eqref{eq:CB_gaussian_L2_factor} cancels the $z_1$-dependent exponential in Eq.~\eqref{eq:CB_theta_L2_factor}.  Therefore the full boundary condition reduces to
\begin{equation}
\begin{aligned}
    &\varphi_{m_1}(z_1+\ii L_2)
    \FCM(Z+\ii L_2)
    \ee^{\ii2\pi(2Z/L_1+\tau)}
    \ee^{\ii\pi\Nd(2x_1/L_1)}\\
    &\qquad=\varphi_{m_1}(z_1)\FCM(Z).
\end{aligned}
    \label{eq:CB_intermediate_L2}
\end{equation}
The reduced-flux single-particle orbital satisfies the corresponding reference boundary condition
\begin{equation}
    \varphi_{m_1}(z_1+\ii L_2)
    \ee^{\ii\pi\Nd(2x_1/L_1)}
    =\varphi_{m_1}(z_1).
    \label{eq:reduced_orbital_L2_BC}
\end{equation}
Inserting Eq.~\eqref{eq:reduced_orbital_L2_BC} into Eq.~\eqref{eq:CB_intermediate_L2} yields the required center-of-mass boundary condition
\begin{equation}
    \boxed{
    \FCM(Z+\ii L_2)
    \ee^{\ii2\pi(2Z/L_1+\tau)}
    =\FCM(Z).}
    \label{eq:CB_FCM_L2}
\end{equation}
Eqs.~\eqref{eq:CB_FCM_L1} and \eqref{eq:CB_FCM_L2} are the boundary conditions used for the reference composite-boson center-of-mass factor in the rectangular geometry.

A convenient way of satisfying these two conditions is to choose a level-two center-of-mass theta function.  For example,
\begin{equation}
    \FCM^{(a)}(Z)=
    \vartheta_{a,0}\!\left(\frac{2Z}{L_1}\middle|2\tau\right),
    \qquad a=0,\frac12,
    \label{eq:example_FCM_solution}
\end{equation}
obeys Eqs.~\eqref{eq:CB_FCM_L1} and \eqref{eq:CB_FCM_L2} (in our numerics we choose \(a=0\)).  Other equivalent bases are related by linear combinations and shifts of the theta-function characteristics.

\subsection{Mapping to the physical Hofstadter convention}
\label{subsec:mapping_to_physical_convention}

The construction above gives the opposite-chirality reference state.
For a reference seed associated with the reduced-flux occupation pattern
\(\mathbf n\), the seed used for the physical positive-flux Hofstadter
Hamiltonian is its complex conjugate,
\begin{equation}
    \Psi_{\mathbf n}^{(0)}(+\phi)
    =\left[
    \Psi_{\mathbf n,{\rm ref}}^{(0)}(-\phi)
    \right]^* .
    \label{eq:reference_to_physical_state}
\end{equation}
The conjugation in Eq.~\eqref{eq:reference_to_physical_state} acts on the
complete trial state, including the reduced-flux orbitals, center-of-mass
factor, and Jastrow factor. Conjugating only the magnetic-translation phase
while retaining the reference wave-function factors would combine
ingredients with opposite chiralities.

At the single-particle level, this conjugation changes the translation
eigenvalue to its complex conjugate. Since a translation along the \(x\)
direction carries no additional gauge phase in either convention,
\begin{equation}
    t_1^{(+)}\!\left(\frac{L_1}{M}\right)
    \left[\varphi_{m,{\rm ref}}^{(M)}\right]^*
    =
    \ee^{-2\pi\ii m/M}
    \left[\varphi_{m,{\rm ref}}^{(M)}\right]^* .
    \label{eq:physical_magnetic_momentum}
\end{equation}
If the physical orbitals are relabelled so that their eigenvalues are
written as \(\ee^{2\pi\ii\bar m/M}\), their magnetic-momentum labels are
therefore
\begin{equation}
    \bar m=-m\pmod M .
    \label{eq:magnetic_momentum_reversal}
\end{equation}
Complex conjugation thus reverses the cyclic ordering of the
magnetic-momentum labels while leaving the orbital densities unchanged.

Returning to the physical Hofstadter flux
\(\phi=\Nphi/(L_1L_2)\), the magnetic translation along the \(y\)
direction is obtained by the same conjugation. In particular, the
\(L_2\)-period translation carries the conjugate gauge phase,
\begin{equation}
    \bigl(t_{2}^{(+)}\psi\bigr)(x,y)
    =\ee^{-\ii2\pi\Nphi x/L_1}\psi(x,y+L_2).
    \label{eq:physical_period_translation}
\end{equation}
To define the center-of-mass translations used below, let
\(t_{2,{\rm ref}}(1)\) and \(t_2^{(+)}(1)\) denote the reference and
physical single-particle translations by one lattice spacing along the
\(y\) direction:
\begin{align}
    \bigl(t_{2,{\rm ref}}(1)\psi\bigr)(x,y)
    &=\ee^{+\ii2\pi\phi x}\psi(x,y+1),\\
    \bigl(t_2^{(+)}(1)\psi\bigr)(x,y)
    &=\ee^{-\ii2\pi\phi x}\psi(x,y+1).
\end{align}
Thus
\begin{equation}
    t_{2,{\rm ref}}=\bigl[t_{2,{\rm ref}}(1)\bigr]^{L_2},
    \qquad
    t_2^{(+)}=\bigl[t_2^{(+)}(1)\bigr]^{L_2}.
\end{equation}
The corresponding many-body center-of-mass translations are
\begin{equation}
    T_{\mathrm{CM,ref}}
    =\prod_{j=1}^{N}\bigl[t_{2,{\rm ref}}(1)\bigr]^{(j)},
    \qquad
    T_{\mathrm{CM}}
    =\prod_{j=1}^{N}\bigl[t_2^{(+)}(1)\bigr]^{(j)},
\end{equation}
where the superscript \((j)\) indicates that the single-particle operator
acts on particle \(j\). Their actions on the reference and physical
seeds are related by
\begin{equation}
    (T_{\mathrm{CM}})\Psi_{\mathbf n}^{(0)}(+\phi)
    =
    \left[
    (T_{\mathrm{CM,ref}})
    \Psi_{\mathbf n,{\rm ref}}^{(0)}(-\phi)
    \right]^* .
    \label{eq:reference_to_physical_translation}
\end{equation}
Equation~\eqref{eq:reference_to_physical_translation} maps every reference
center-of-mass translation orbit to a corresponding physical orbit through
complex conjugation. The translation eigenphases are conjugated, thereby
relabelling the many-body magnetic-translation sectors, while orbit sizes,
relations among trial states, state counting, and trial-space rank remain
unchanged. Together with
\(H(+\phi)=[H(-\phi)]^*\), which holds for the HBH Hamiltonian, this also
gives identical reference and physical energy spectra.

\section{Counting sketch for the reduced ansatz degeneracy}
\label{sec:counting-reduction-sketch}

This section gives a simple counting interpretation of the degeneracies observed in the composite-boson ansatz. Here we explain why the ansatz naturally organizes the naive reduced-flux configurations into cyclic orbits, and why this gives the counting reduction seen in the variational construction. We first describe the cyclic shift in the reference convention; complex conjugation maps it to the physical convention without changing the orbit decomposition or the count.

A recent complementary construction used an overcomplete set of
position-localized parton quasihole states and momentum-resolved Gram matrices
to recover the established two- and three-quasihole counting in generalized
Kapit--Mueller and checkerboard models; in the checkerboard model, the
resulting states also showed high overlap with the exact low-energy
subspace~\cite{LotricSimon2026}. The construction tested here instead uses
unpinned bosonic occupations of the reduced-flux orbitals and resolves their
cyclic-translation redundancy orbit by orbit.

We denote the number of particles by $N$ and the total number of flux quanta by $\Nphi$. After attaching two flux quanta to each boson, the reduced flux is
\begin{equation}
    \Nd=\Nphi-2N .
\end{equation}
The number of bosonic occupation patterns of $N$ particles in $\Nd$ reduced-flux orbitals is
\begin{equation}
    C_{\rm B}=\binom{\Nd+N-1}{N} .
\end{equation}
We also define
\begin{equation}
    g=\gcd(N,\Nd)=\gcd(N,\Nphi),
    \qquad
    n=\frac{\Nd}{g},
    \qquad
    \qcom=\frac{\Nphi}{g} .
\end{equation}
Here $\qcom$ is the center-of-mass degeneracy associated with magnetic translations. The integer $n=\Nd/g$ is the reduction factor appearing in the occupation-orbit counting below.

\subsection{Fixed center-of-mass factor and full center-of-mass partners}
\label{subsec:cm-factor}

By fixing the center-of-mass sector in the seed ansatz, we mean that the seed
states are built with one chosen reference center-of-mass theta factor
$\FCM^{(0)}(Z)$ satisfying the boundary conditions in
Eqs.~\eqref{eq:CB_FCM_L1} and \eqref{eq:CB_FCM_L2}.

The remaining reference center-of-mass partners are generated afterward by the
many-body magnetic translation along the $y$ direction. In the reference
convention used above, a translation by $m$ lattice spacings acts on a
many-body wave function as
\begin{equation}
    \bigl((T_{\mathrm{CM,ref}})^m\Psi_{\rm ref}\bigr)(\{z_j\})
    =
    \ee^{\ii 2\pi\phi m\sum_{j=1}^N x_j}
    \Psi_{\rm ref}(\{z_j+\ii m\}) .
    \label{eq:TCM_action_general}
\end{equation}
To apply this to the seed, we write
\begin{equation}
    \Psi_{\mathrm{seed,ref}}(\{z_j\})
    =
    \Phi_{\mathrm{seed,ref}}(\{z_j\})
    \FCM^{(0)}(Z)
    \exp\left[
    -\frac{\pi\Nphi}{L_1L_2}
    \sum_{j=1}^N y_j^2
    \right],
    \label{eq:seed_factored_for_TCM}
\end{equation}
where $\Phi_{\mathrm{seed,ref}}$ contains the reduced-flux orbital part and the
Jastrow factor, with the final physical Gaussian factored out. Since the
Jastrow factor depends only on coordinate differences, it is invariant under a
common translation of all particles. The magnetic phase in
Eq.~\eqref{eq:TCM_action_general} combines with the linear term generated by
translating the Gaussian, and one obtains
\begin{equation}
\begin{split}
    \bigl((T_{\mathrm{CM,ref}})^m\Psi_{\mathrm{seed,ref}}\bigr)(\{z_j\})
    ={}&
    \ee^{-\pi\Nphi N m^2/(L_1L_2)}
    \ee^{\ii 2\pi\Nphi m Z/(L_1L_2)}
    \Phi_{\mathrm{seed,ref}}(\{z_j+\ii m\})
    \FCM^{(0)}(Z+\ii mN)
    \\
    &\times
    \exp\left[
    -\frac{\pi\Nphi}{L_1L_2}
    \sum_{j=1}^N y_j^2
    \right] .
\end{split}
    \label{eq:TCM_action_seed}
\end{equation}
For the square choice $L_1=L_2=\Nphi$, the two prefactors in
Eq.~\eqref{eq:TCM_action_seed} reduce to \(\ee^{-\pi N m^2/\Nphi}\,\ee^{\ii 2\pi m Z/\Nphi}\).

The physical center-of-mass translation carries the conjugate phase,
\begin{equation}
    \bigl((T_{\mathrm{CM}})^m\Psi\bigr)(\{x_j,y_j\})
    =
    \ee^{-\ii 2\pi\phi m\sum_{j=1}^N x_j}
    \Psi(\{x_j,y_j+m\}) ,
    \label{eq:TCM_action_physical}
\end{equation}
and its action on the physical seed is the complex conjugate of the reference
action in Eq.~\eqref{eq:TCM_action_seed}. Complex conjugation reverses the physical magnetic-momentum labeling without
changing the multiplet size.

The set
\(\left\{(T_{\mathrm{CM,ref}})^m\Psi_{\mathrm{seed,ref}}\right\}_{m=0}^{\qcom-1}\)
generates the full $\qcom$-fold reference center-of-mass multiplet, and its
complex conjugate gives the corresponding physical multiplet. This is where
the center-of-mass multiplicity enters the ansatz counting.

\subsection{Real-space translations and cyclic occupation orbits}
\label{subsec:translation-to-orbits}

The reduced-flux single-particle reference orbitals of
Eq.~\eqref{eq:single_particle_torus_orbitals} may be written, up to the common Gaussian factor, as
\begin{equation}
    \varphi_{m,{\rm ref}}(z)
    =
    \vartheta_{m/\Nd,0}(u|\tau'),
    \qquad
    u=\frac{\Nd z}{L_1},
    \qquad
    \tau'=\Nd\tau,
    \qquad
    m=0,1,\ldots,\Nd-1 .
\end{equation}
We use the theta-function identity
\begin{equation}
    \vartheta_{a+\delta,0}(u|\tau')
    =
    \ee^{\ii\pi\delta^2\tau'+\ii2\pi\delta u}
    \vartheta_{a,0}(u+\delta\tau'|\tau').
\end{equation}
With $a=m/\Nd$, $\delta=1/\Nd$, and $\tau'=\Nd\tau$, this gives
\begin{equation}
    \varphi_{m,{\rm ref}}(z+\Delta)
    = e^{-i\pi\tau/\Nd}e^{-i2\pi z/L_1}\,\varphi_{m+1,{\rm ref}}(z),
    \qquad
    \Delta=\frac{L_1}{\Nd}\tau,
\end{equation}
where $m+1$ is understood modulo $\Nd$. Thus a real-space translation shifts the reduced-flux orbital label by one, apart from a known prefactor. Here \(\Delta\) is a continuous torus displacement and need not correspond to
an integer number of lattice spacings. We therefore use the cyclic-orbit
construction as a rank prediction for the trial family generated by discrete
one-site translations and test this prediction directly through the
orbit-resolved numerical ranks below.

For a many-body product, translating all particles by the same amount gives
\begin{equation}
    \prod_{j=1}^N \varphi_{m_j,{\rm ref}}(z_j+\Delta)
    = e^{-i\pi N\tau/\Nd}e^{-i2\pi Z/L_1}
      \prod_{j=1}^N \varphi_{m_j+1,{\rm ref}}(z_j),
    \qquad Z=\sum_{j=1}^N z_j .
\end{equation}
The extra factor depends only on the center-of-mass coordinate $Z$. For the occupation counting, the important point is simply that all reduced-flux orbital labels are shifted by one. Therefore a bosonic occupation pattern
\begin{equation}
    {\bf n}=(n_0,n_1,\ldots,n_{\Nd-1}),
    \qquad \sum_{r=0}^{\Nd-1} n_r=N,
\end{equation}
is mapped, up to common factors, to the cyclically shifted pattern
\begin{equation}
    \Tcyc:
    (n_0,n_1,\ldots,n_{\Nd-1})
    \mapsto
    (n_{\Nd-1},n_0,\ldots,n_{\Nd-2}) .
\end{equation}
This is the basic reason why the naive $C_{\rm B}$ configurations should be organized into cyclic orbits. Under the physical relabelling
\(\bar m=-m\pmod{\Nd}\) defined above, the reference shift
\(m\mapsto m+1\) becomes \(\bar m\mapsto\bar m-1\).  It is therefore
represented by \(\Tcyc^{-1}\) when the physical orbitals are labelled by
\(\bar m\); if they retain the reference labels \(m\), it remains
\(\Tcyc\). The cyclic orbits and their sizes are identical in either labeling. The Jastrow factor depends only on coordinate differences and does not change this occupation-space bookkeeping. The center-of-mass theta factor enforces the reference magnetic boundary conditions, and its conjugate enforces the physical ones, but neither alters the cyclic permutation of the reduced-flux occupation patterns.

\subsection{Orbit sizes and the reduction factor}
\label{subsec:orbit-size-reduction}

Let $s$ be the size of one cyclic orbit under $\Tcyc$. If an occupation
pattern repeats after $s$ cyclic shifts, then the $\Nd$ orbitals consist of
\begin{equation}
    h=\frac{\Nd}{s}
\end{equation}
identical blocks. Since the same number of particles must occupy each block,
the total particle number can be written as
\begin{equation}
    N=hN_{\rm block}
\end{equation}
for some integer $N_{\rm block}$. Hence $h$ divides $N$. By definition,
$h=\Nd/s$ also divides $\Nd$, and therefore $h$ divides $\gcd(N,\Nd)$. We may thus write
\begin{equation}
    g=hw
\end{equation}
with integer $w$. Using $s=\Nd/h$, it follows that
\begin{equation}
    s=\frac{\Nd}{h}
     =\frac{\Nd}{g}\frac{g}{h}
     =\frac{\Nd}{g}\,w .
\end{equation}
Therefore every orbit size is a multiple of
\begin{equation}
    n=\frac{\Nd}{g},
\end{equation}
so that
\begin{equation}
    s=nw
\end{equation}
with integer $w$.

Now take one orbit and label its occupation patterns by
\begin{equation}
    |0\rangle,|1\rangle,\ldots,|s-1\rangle,
\end{equation}
where $|t\rangle$ is the $t$th cyclic shift of a representative occupation pattern. Since $s=nw$, the same labels may be written as
\begin{equation}
    t=\mu+aw,
    \qquad
    \mu=0,1,\ldots,w-1,
    \qquad
    a=0,1,\ldots,n-1 .
\end{equation}
The $s$ labels are thereby partitioned into the $w=s/n$ disjoint strings
\begin{equation}
    \mathcal S_{\mu}
    =\{\mu,\mu+w,\ldots,\mu+(n-1)w\},
    \qquad \mu=0,1,\ldots,w-1 .
\end{equation}
The \(n\) configurations in a given string are related by repeated
applications of \(\Tcyc^w\). Since \(\Tcyc^w\) shifts all reduced-flux orbital labels by the same amount
while preserving their relative occupation pattern, the members of a string
are naturally viewed as translated versions of the same internal
configuration. Once the full center-of-mass
multiplet is included, counting all \(n\) members independently would count
the same translated content repeatedly. The reduction is therefore expected
to arise from linear dependencies among the corresponding trial families, so
that only one independent combination per string supplies a new internal
direction. Thus an occupation orbit of size \(s\) predicts
\begin{equation}
    d_O^{\rm int}=w=\frac{s}{n}
\end{equation}
independent internal directions. For example, a size-\(4\) orbit for \((2,8)\) separates into two related
pairs, whereas a size-\(6\) orbit for \((2,10)\) separates into two related
triples; both therefore predict two internal directions. Multiplication by the Jastrow,
center-of-mass, and Gaussian factors completes the variational wave functions;
the predicted rank is tested directly below.

\begin{table}[!htbp]
\centering
\caption{Degeneracy counting for the composite-boson ansatz. Here $\Nd=\Nphi-2N$, $C_{\rm B}=\binom{\Nd+N-1}{N}$, $n=\Nd/g$, $g=\gcd(N,\Nd)$, and $\qcom=\Nphi/g$. The formula column displays the total ansatz count as $\Dtot=(C_{\rm B}/n)\qcom$. The last column lists representative occupation patterns modulo cyclic shifts; the number after the colon is the cyclic-orbit size. Underlined orbit sizes mark the smallest orbit size $s=n$ in that row. The subgroup column records the energy-grouping pattern of the exact
lowest-band-projected spectrum. These energy groups are not identical to the occupation-orbit decomposition, but they carry the signature of the center-of-mass degeneracy: each group size is a multiple of $\qcom$.}
\label{tab:degeneracy-sketch}
\scriptsize
\renewcommand{\arraystretch}{1.25}
\resizebox{\textwidth}{!}{%
\begin{tabular}{@{}c c c c c l l p{0.48\linewidth}@{}}
\toprule
\toprule
$N$ & $\Nphi$ & $\Nd$ & $\nu$ & $n$ & $\Dtot=(C_{\rm B}/n)\qcom$ & Exact-projected subgroups & Inequivalent occupation representatives \\
\midrule
2 & 5  & 1 & $2/5$  & 1 & $(1/1)\times 5=5$   & $5$ & \occ{2}{\uorb{1}} \\
2 & 6  & 2 & $1/3$  & 1 & $(3/1)\times 3=9$   & $6+3$ & \occ{2,0}{2}; \occ{1,1}{\uorb{1}} \\
2 & 7  & 3 & $2/7$  & 3 & $(6/3)\times 7=14$  & $7+7$ & \occ{2,0,0}{\uorb{3}}; \occ{1,1,0}{\uorb{3}} \\
2 & 8  & 4 & $1/4$  & 2 & $(10/2)\times 4=20$ & $4+4+8+4$ & \occ{2,0,0,0}{4}; \occ{1,1,0,0}{4}; \occ{1,0,1,0}{\uorb{2}} \\
2 & 9  & 5 & $2/9$  & 5 & $(15/5)\times 9=27$ & $9+9+9$ & \occ{2,0,0,0,0}{\uorb{5}}; \occ{1,0,1,0,0}{\uorb{5}}; \occ{1,1,0,0,0}{\uorb{5}} \\
2 & 10 & 6 & $1/5$  & 3 & $(21/3)\times 5=35$ & $10+5+5+10+5$ & \occ{2,0,0,0,0,0}{6}; \occ{1,1,0,0,0,0}{6}; \occ{1,0,1,0,0,0}{6}; \occ{1,0,0,1,0,0}{\uorb{3}} \\
2 & 11 & 7 & $2/11$ & 7 & $(28/7)\times 11=44$ & $11+11+11+11$ & \occ{2,0,0,0,0,0,0}{\uorb{7}}; \occ{1,0,0,1,0,0,0}{\uorb{7}}; \occ{1,0,1,0,0,0,0}{\uorb{7}}; \occ{1,1,0,0,0,0,0}{\uorb{7}} \\
2 & 12 & 8 & $1/6$  & 4 & $(36/4)\times 6=54$ & $6+6+12+6+6+12+6$ & \occ{2,0,0,0,0,0,0,0}{8}; \occ{1,0,0,0,1,0,0,0}{\uorb{4}}; \occ{1,0,0,1,0,0,0,0}{8}; \occ{1,0,1,0,0,0,0,0}{8}; \occ{1,1,0,0,0,0,0,0}{8} \\
\midrule
3 & 7  & 1 & $3/7$  & 1 & $(1/1)\times 7=7$   & $7$ & \occ{3}{\uorb{1}} \\
3 & 8  & 2 & $3/8$  & 2 & $(4/2)\times 8=16$  & $8+8$ & \occ{3,0}{\uorb{2}}; \occ{2,1}{\uorb{2}} \\
3 & 9  & 3 & $1/3$  & 1 & $(10/1)\times 3=30$ & $12+3+3+12$ & \occ{3,0,0}{3}; \occ{2,1,0}{3}; \occ{2,0,1}{3}; \occ{1,1,1}{\uorb{1}} \\
3 & 10 & 4 & $3/10$ & 4 & $(20/4)\times 10=50$ & $10+10+10+10+10$ & \occ{3,0,0,0}{\uorb{4}}; \occ{2,1,0,0}{\uorb{4}}; \occ{2,0,1,0}{\uorb{4}}; \occ{2,0,0,1}{\uorb{4}}; \occ{1,1,1,0}{\uorb{4}} \\
\midrule
4 & 9  & 1 & $4/9$  & 1 & $(1/1)\times 9=9$   & 9 & \occ{4}{\uorb{1}} \\
4 & 10 & 2 & $2/5$  & 1 & $(5/1)\times 5=25$  & $5+10+5+5$ & \occ{4,0}{2}; \occ{3,1}{2}; \occ{2,2}{\uorb{1}} \\
\bottomrule
\bottomrule
\end{tabular}%
}
\end{table}

Summing these predicted orbit contributions gives
\begin{equation}
    \Dint
    =\sum_{\rm orbits}\frac{s}{n}
    =\frac{1}{n}\sum_{\rm orbits}s
    =\frac{C_{\rm B}}{n}
    =C_{\rm B}\frac{g}{\Nd} .
\end{equation}
Including the $\qcom$ center-of-mass partners gives the total ansatz count
\begin{equation}
    \Dtot=\Dint\qcom
    =\frac{C_{\rm B}}{n}\,\qcom
    =C_{\rm B}\frac{\Nphi}{\Nd} .
\label{eq:degeneracy-count}
\end{equation}
Table~\ref{tab:degeneracy-sketch} summarizes this counting for a broader set
of systems and also lists the energy-subgroup multiplicities found
numerically. The energy subgroups and the occupation-orbit blocks are distinct
structures, even when their multiplicities happen to coincide.

\subsection{Orbit-resolved numerical rank test}
\label{subsec:orbit-rank-test}

We test the reduction directly at the level of the constructed wave functions,
without using the Hamiltonian. Let
$O=\{\mathbf n_{O,t}\}_{t=0}^{s_O-1}$ be a cyclic orbit of reduced-flux
occupation patterns. From each pattern $\mathbf n_{O,t}$ we construct all
$\qcom$ center-of-mass partners $|\Phi_{O,t,m}\rangle$, with
$m=0,\ldots,\qcom-1$. In a fixed real-space many-body configuration basis
$\{|C_\alpha\rangle\}$, these states are collected as the columns of
\begin{equation}
    (A_O)_{\alpha,(t,m)}
    =\langle C_\alpha|\Phi_{O,t,m}\rangle .
    \label{eq:orbit-state-block}
\end{equation}
Thus $A_O$ has $s_O\qcom$ columns, and its rank is the number of linearly
independent trial states contributed by orbit $O$. Writing its singular values
in descending order, $\sigma_1(A_O)\geq\sigma_2(A_O)\geq\cdots$, we count
$\sigma_k(A_O)$ as nonzero when
\begin{equation}
    \frac{\sigma_k(A_O)}{\sigma_1(A_O)}>10^{-10} .
    \label{eq:orbit-rank-cutoff}
\end{equation}
Here $\sigma_1(A_O)$ is the largest singular value of the same orbit block, so
the criterion is independent of its overall normalization and treats as
numerically null only directions whose singular value is at least ten orders
of magnitude smaller than the leading one. Equivalently,
$A_O^\dagger A_O$ is the overlap matrix of these $s_O\qcom$ trial states and
has eigenvalues $\lambda_k=\sigma_k^2(A_O)$. Thus
Eq.~\eqref{eq:orbit-rank-cutoff} corresponds to
$\lambda_k/\lambda_1>10^{-20}$ for that overlap matrix. In practice, we work
directly with the singular values of $A_O$, which avoids squaring the numerical
range.

\begin{table}[!htbp]
\centering
\caption{Orbit-resolved rank test of the cyclic reduction. Each parameter line
gives the quantities entering $n=\Nd/g$, with $g=\gcd(N,\Nd)$, together with
the naive occupation count $C_{\rm B}$, the total number
$N_{\rm raw}=C_{\rm B}\qcom$ of raw translated states, and the full rank
$\mathcal D$. Here $\mathbf n_O$ is one representative reduced-flux occupation
pattern from orbit $O$, and $s_O$ is the size of that orbit. For each orbit,
$N_O^{\rm raw}=s_O\qcom$ and $r_O=(s_O/n)\qcom$. The listed $r_O$ is both the
predicted and the independently obtained numerical rank; the two agree in
every row.}
\label{tab:orbit-rank-test}
\small
\setlength{\tabcolsep}{7pt}
\renewcommand{\arraystretch}{1.12}
\begin{tabular}{@{}l c c c@{}}
\toprule
\toprule
\multicolumn{4}{@{}l}{\((N,\Nphi)=(2,8):\ \Nd=4,\ g=2,\ n=2,\ \qcom=4,\ C_{\rm B}=10,\ N_{\rm raw}=40,\ \mathcal D=20\)} \\
$\mathbf n_O$ & $s_O$ & $N_O^{\rm raw}$ & $r_O^{\rm pred}=r_O^{\rm num}$ \\
\midrule
$(2,0,0,0)$ & 4 & 16 & 8 \\
$(1,1,0,0)$ & 4 & 16 & 8 \\
$(1,0,1,0)$ & 2 & 8 & 4 \\
\midrule
\midrule
\multicolumn{4}{@{}l}{\((N,\Nphi)=(2,10):\ \Nd=6,\ g=2,\ n=3,\ \qcom=5,\ C_{\rm B}=21,\ N_{\rm raw}=105,\ \mathcal D=35\)} \\
$\mathbf n_O$ & $s_O$ & $N_O^{\rm raw}$ & $r_O^{\rm pred}=r_O^{\rm num}$ \\
\midrule
$(2,0,0,0,0,0)$ & 6 & 30 & 10 \\
$(1,1,0,0,0,0)$ & 6 & 30 & 10 \\
$(1,0,1,0,0,0)$ & 6 & 30 & 10 \\
$(1,0,0,1,0,0)$ & 3 & 15 & 5 \\
\midrule
\midrule
\multicolumn{4}{@{}l}{\((N,\Nphi)=(3,9):\ \Nd=3,\ g=3,\ n=1,\ \qcom=3,\ C_{\rm B}=10,\ N_{\rm raw}=30,\ \mathcal D=30\)} \\
$\mathbf n_O$ & $s_O$ & $N_O^{\rm raw}$ & $r_O^{\rm pred}=r_O^{\rm num}$ \\
\midrule
$(3,0,0)$ & 3 & 9 & 9 \\
$(2,1,0)$ & 3 & 9 & 9 \\
$(2,0,1)$ & 3 & 9 & 9 \\
$(1,1,1)$ & 1 & 3 & 3 \\
\midrule
\midrule
\multicolumn{4}{@{}l}{\((N,\Nphi)=(3,10):\ \Nd=4,\ g=1,\ n=4,\ \qcom=10,\ C_{\rm B}=20,\ N_{\rm raw}=200,\ \mathcal D=50\)} \\
$\mathbf n_O$ & $s_O$ & $N_O^{\rm raw}$ & $r_O^{\rm pred}=r_O^{\rm num}$ \\
\midrule
$(3,0,0,0)$ & 4 & 40 & 10 \\
$(2,1,0,0)$ & 4 & 40 & 10 \\
$(2,0,1,0)$ & 4 & 40 & 10 \\
$(2,0,0,1)$ & 4 & 40 & 10 \\
$(1,1,1,0)$ & 4 & 40 & 10 \\
\bottomrule
\bottomrule
\end{tabular}
\end{table}

An orbit of size $s_O$ contains
\begin{equation}
    N_O^{\rm raw}=s_O\qcom
\end{equation}
raw translated states, while the cyclic reduction predicts
\begin{equation}
    r_O=\frac{s_O}{n}\qcom .
    \label{eq:orbit-rank-prediction}
\end{equation}
Table~\ref{tab:orbit-rank-test} gives the orbit-resolved comparison for the
four systems whose ansatz-subspace spectra are shown in the main text. In
every orbit, the independently computed numerical rank equals
Eq.~\eqref{eq:orbit-rank-prediction}; the table therefore displays a single
rank column. Moreover, the orbit ranks add to the full numerical rank in each
system, so no additional rank reduction occurs between distinct orbit blocks.
This agreement is substantially more restrictive than reproducing only the
total dimension $\mathcal D$: it verifies the predicted linear-dependence
pattern separately within every cyclic occupation orbit, including shortened
orbits and reduction factors $n=1,2,3,$ and $4$. The orbit-by-orbit match thus
provides direct numerical support for the reduced-flux bosonic organization
underlying the composite-boson ansatz.

\section{Evaluation of the many-body Chern number}
\label{app:many-body-chern}

We compute the many-body Chern number of a selected low-energy manifold by using the gauge-invariant lattice formulation of the Berry curvature in the space of boundary twists~\cite{FukuiHatsugaiSuzuki2005}.  Let
\begin{equation}
    \bm{\theta}=(\theta_x,\theta_y)
\end{equation}
be the two twist angles inserted through the two non-contractible cycles of the torus.  In practice, this parameter space is replaced by a uniform mesh,
\begin{equation}
    \theta_i = \frac{2\pi}{N_\theta} n_i,
    \qquad n_i=0,1,\ldots,N_\theta-1,
    \qquad i=x,y,
\end{equation}
where the integer $N_\theta$ fixes the resolution of the discretization.  The elementary displacements on this mesh are denoted by
\begin{equation}
    \bm{\delta}_x=\left(\frac{2\pi}{N_\theta},0\right),
    \qquad
    \bm{\delta}_y=\left(0,\frac{2\pi}{N_\theta}\right).
\end{equation}
The twist angles implement generalized, or twisted, boundary conditions; equivalently, they correspond to fluxes threaded through the handles of the torus~\cite{NiuThoulessWu1985}.

At each point \(\bm{\theta}\), we diagonalize the many-body Hamiltonian
with the corresponding boundary twists. Suppose that the Chern number is
to be assigned to a \(\mathcal D\)-dimensional unpinned manifold, such as
the \(\mathcal D\) lowest states separated from the rest of the spectrum
by a gap. We collect an orthonormal basis of this manifold into the
multiplet matrix
\begin{equation}
    \Phi(\bm{\theta})
    =\bigl(
    |G_1(\bm{\theta})\rangle,
    |G_2(\bm{\theta})\rangle,
    \ldots,
    |G_{\mathcal D}(\bm{\theta})\rangle
    \bigr).
\end{equation}

The link variable associated with a step in the \(i=x,y\) direction is
defined from the determinant of the overlap matrix between neighboring
multiplets,
\begin{equation}
    U_i(\bm{\theta})
    =
    \frac{
    \det\!\left[
    \Phi^\dagger(\bm{\theta})
    \Phi(\bm{\theta}+\bm{\delta}_i)
    \right]
    }{
    \left|
    \det\!\left[
    \Phi^\dagger(\bm{\theta})
    \Phi(\bm{\theta}+\bm{\delta}_i)
    \right]
    \right|},
    \qquad i=x,y .
    \label{eq:link_variable}
\end{equation}

Here the overlap is taken between physical many-body states. To make this
explicit, write
\[
    |G_\alpha(\bm{\theta})\rangle
    =
    \sum_{\mathbf n}
    c_{\mathbf n\alpha}(\bm{\theta})
    |\mathbf n;\bm{\theta}\rangle,
\]
where \(|\mathbf n;\bm{\theta}\rangle\) denotes a many-body occupation state
constructed from the lowest-band orbitals at \(\bm{\theta}\). For
\(\bm{\theta}'=\bm{\theta}+\bm{\delta}_i\), the overlap matrix entering
Eq.~\eqref{eq:link_variable} is therefore
\[
    \bigl[\Phi^\dagger(\bm{\theta})\Phi(\bm{\theta}')\bigr]_{\alpha\beta}
    =
    \sum_{\mathbf n,\mathbf m}
    c_{\mathbf n\alpha}^{*}(\bm{\theta})\,
    B_{\mathbf n\mathbf m}(\bm{\theta},\bm{\theta}')\,
    c_{\mathbf m\beta}(\bm{\theta}'),
\]
where
\[
    B_{\mathbf n\mathbf m}(\bm{\theta},\bm{\theta}')
    =
    \langle\mathbf n;\bm{\theta}
    \mid\mathbf m;\bm{\theta}'\rangle .
\]
Equivalently, if \(C(\bm{\theta})\) collects the coefficients
\(c_{\mathbf n\alpha}(\bm{\theta})\), the overlap matrix is
\(C^\dagger(\bm{\theta})B(\bm{\theta},\bm{\theta}')C(\bm{\theta}')\).
In a common site-occupation basis (B) is the identity. In the projected calculation, its matrix elements are instead obtained as normalized permanents of matrices built from overlaps between the lowest-band orbitals at the two twist points.

The link variable \(U_i\) retains only the phase of the determinant. For
\(\mathcal D=1\), it reduces to the usual Abelian Berry link. For
\(\mathcal D>1\), the determinant construction treats the entire multiplet
at once, and the resulting plaquette phase is insensitive to arbitrary
unitary rotations among the states within the chosen subspace. The Berry
curvature assigned to the plaquette whose lower-left corner is
\(\bm{\theta}\) is then
\begin{equation}
    F(\bm{\theta})
    =
    \log\!\left[
    U_x(\bm{\theta})
    U_y(\bm{\theta}+\bm{\delta}_x)
    U_x(\bm{\theta}+\bm{\delta}_y)^{-1}
    U_y(\bm{\theta})^{-1}
    \right],
    \label{eq:discrete_curvature}
\end{equation}
where the logarithm is taken on the principal branch.  The many-body Chern number of the chosen manifold follows by summing this discrete curvature over the full twist-angle Brillouin zone,
\begin{equation}
    C_{\rm MB}
    =
    \frac{1}{2\pi i}
    \sum_{\bm{\theta}} F(\bm{\theta}) .
    \label{eq:chern_number}
\end{equation}
In numerical calculations, the mesh should be refined until the integer value of $C_{\rm MB}$ is stable.  The selected multiplet should also remain separated from higher states throughout the twist-angle mesh, so that the same $\mathcal D$-dimensional subspace is followed over the entire parameter space.

Table~\ref{tab:chern_counting} summarizes our numerical many-body
Chern-number calculations for unpinned systems. The results reported here were obtained using an
\(N_\theta\times N_\theta\) mesh with \(N_\theta=21\). In every case, the
manifold-averaged Chern number per state satisfies
\(C_{\rm MB}/\mathcal D=N/N_\phi=\nu\); in the presence of localized flux, the analogous
relation \(C_{\rm MB}/D=\nu_{\mathrm{eff}}
=N/(N_\phi-N_{\mathrm{loc}})\) was established in our earlier
work~\cite{ZeybekUmucalilar2022}.

\begin{table}[t]
\caption{\label{tab:chern_counting}
Dimension \(\mathcal D\) and total many-body Chern number \(C_{\rm MB}\) of the
unpinned low-energy manifold for \(N_d=N_\phi-2N>0\).
For every listed system, the numerically obtained values of both
\(\mathcal D\) and \(C_{\rm MB}\) agree to numerical precision with the combinatorial predictions
\(\mathcal{D}=\binom{N_d+N-1}{N}N_\phi/N_d\)
[Eq.~\eqref{eq:degeneracy-count}] and
\(C_{\rm MB}=\binom{N_d+N-1}{N_d}\), respectively.
Consequently, the manifold-averaged Chern number per state is
\(C_{\rm MB}/\mathcal D=N/N_\phi=\nu\).}
\centering
\small
\setlength{\tabcolsep}{4.5pt}
\renewcommand{\arraystretch}{1.08}
\begin{ruledtabular}
\begin{tabular}{cccccc}
\(N\) & \(N_\phi\) & \(N_d\) & \(\mathcal D\) & \(C_{\rm MB}\) & \(C_{\rm MB}/\mathcal D=\nu\) \\
\hline
2 &  5 & 1 &   5 &  2 & \(2/5\)  \\
2 &  6 & 2 &   9 &  3 & \(1/3\)  \\
2 &  7 & 3 &  14 &  4 & \(2/7\)  \\
2 &  8 & 4 &  20 &  5 & \(1/4\)  \\
2 &  9 & 5 &  27 &  6 & \(2/9\)  \\
2 & 10 & 6 &  35 &  7 & \(1/5\)  \\
2 & 11 & 7 &  44 &  8 & \(2/11\) \\
\hline
3 &  7 & 1 &   7 &  3 & \(3/7\)  \\
3 &  8 & 2 &  16 &  6 & \(3/8\)  \\
3 &  9 & 3 &  30 & 10 & \(1/3\)  \\
3 & 10 & 4 &  50 & 15 & \(3/10\) \\
\hline
4 &  9 & 1 &   9 &  4 & \(4/9\)  \\
4 & 10 & 2 &  25 & 10 & \(2/5\)  \\
4 & 12 & 4 & 105 & 35 & \(1/3\)  \\
\hline
5 & 11 & 1 &  11 &  5 & \(5/11\) \\
\end{tabular}
\end{ruledtabular}
\end{table}

\section{Numerical Berry phase for braiding pinned quasiholes}

We calculate the Berry phase by adiabatically moving the pinning potentials along a closed path and following the low-energy quasihole manifold. Here and below, ``quasihole'' denotes the localized flux defect created by
the pin within the selected sub-half-filled manifold. Since this manifold is nearly degenerate, the result is in general a Berry matrix rather than a single phase~\cite{Berry1984,WilczekZee1984}.  The numerical procedure is a discrete parallel-transport construction for a degenerate subspace.  It is the same basic idea used in multiband Berry-phase and Wilson-loop calculations, where the states at neighboring points are aligned by a singular-value decomposition of their overlap matrix~\cite{Vanderbilt2018}.  A closely related approach was used in numerical braiding calculations for lattice bosons with moving local potentials~\cite{KapitGinspargMueller2011}.

Let the pinning potentials define a closed path \(\CC\) in parameter space, and write the Hamiltonian along this path as
\begin{equation}
  H(\lambda)=H_0+H_{\mathrm{pin}}(\lambda).
  \label{eq:Hlambda}
\end{equation}
Here \(H_0\) is the unpinned many-body Hamiltonian, and \(H_{\mathrm{pin}}\) contains the local potentials used to pin and move the quasiholes.  At each point \(\lambda\), we keep the lowest \(D\) states, assuming that they form the quasihole manifold of interest and remain separated from higher states.  These states are collected as columns of the matrix
\begin{equation}
  \Psi(\lambda)=\bigl(\ket{\psi_1(\lambda)},\ldots,
  \ket{\psi_D(\lambda)}\bigr),
  \qquad
  \Psi^{\dagger}(\lambda)\Psi(\lambda)=\mathbbm{1}_{D}.
  \label{eq:Psimatrix}
\end{equation}
In the finite systems studied here the manifold is usually only nearly
degenerate.  We isolate the geometric part by parallel-transporting the
low-energy subspace without accumulating dynamical phases from its small
internal splittings, while separately monitoring those splittings and the gap
above the manifold.  The geometric Berry matrix is defined operationally from
the endpoint mismatch after this discrete transport, as specified below.

\subsection{Discrete parallel transport}

The path is sampled by a sequence of nearby Hamiltonians \(H(\lambda_0),H(\lambda_1),\ldots,H(\lambda_L)\), with \(\lambda_L\) equivalent to \(\lambda_0\).  We use the cross-shaped commutator loop
\begin{equation}
  \CC=T_2^{-1}T_1^{-1}T_2T_1,
  \label{eq:commutator_loop}
\end{equation}
whose four transport segments, read in their physical order, are
\(T_1,T_2,T_1^{-1},T_2^{-1}\). Here, \(T_1\) and \(T_2\) label the quasihole-transport segments and are
unrelated to the magnetic-translation operators defined previously. One quasihole is held fixed during each
segment while the other is transported.  The oppositely directed transport segments are paired so that their Aharonov--Bohm contributions cancel, isolating the mutual braiding contribution
to the geometric phase~\cite{WenDagottoFradkin1990SM}. In practice, each elementary displacement is implemented by transferring the
pinning potential linearly between neighboring lattice sites, lowering its
strength at the departure site while raising it at the destination site.

At a new point \(\lambda_\ell\), diagonalization gives a raw orthonormal basis for the selected manifold,
\begin{equation}
  \widetilde{\Psi}_{\ell}
  =\bigl(\ket{\widetilde{\psi}_{1,\ell}},\ldots,
  \ket{\widetilde{\psi}_{D,\ell}}\bigr).
  \label{eq:rawbasis}
\end{equation}
This basis is not yet smooth along the path.  The phases of individual eigenvectors are arbitrary, and within a nearly degenerate subspace the diagonalization may also choose an arbitrary unitary rotation of the basis.  To remove this ambiguity, we align the new basis with the transported basis from the previous step.

Suppose \(\Psi_{\ell-1}\) has already been fixed.  We form the overlap matrix
\begin{equation}
  A_{\ell}=\Psi_{\ell-1}^{\dagger}\widetilde{\Psi}_{\ell}
  \label{eq:overlapA}
\end{equation}
and compute its singular value decomposition,
\begin{equation}
  A_{\ell}=V_{\ell}\Sigma_{\ell}W_{\ell}^{\dagger}.
  \label{eq:SVD}
\end{equation}
The unitary part of this overlap is
\begin{equation}
  Q_{\ell}=V_{\ell}W_{\ell}^{\dagger}.
  \label{eq:polar}
\end{equation}
We then define the transported basis at the new point by
\begin{equation}
  \Psi_{\ell}=\widetilde{\Psi}_{\ell}Q_{\ell}^{\dagger}.
  \label{eq:transported_basis}
\end{equation}
With this choice,
\begin{equation}
  \Psi_{\ell-1}^{\dagger}\Psi_{\ell}
  =A_{\ell}Q_{\ell}^{\dagger}
  =V_{\ell}\Sigma_{\ell}V_{\ell}^{\dagger},
  \label{eq:positive_overlap}
\end{equation}
which is Hermitian and positive.  Thus the basis at \(\lambda_\ell\) is chosen to be as close as possible to the basis at \(\lambda_{\ell-1}\).  For a single state this reduces to the familiar rule of choosing the phase of the new state so that its overlap with the previous state is real and positive.  For several states, it gives the corresponding optimal alignment of the whole subspace.

A useful check of the smoothness of the path is
\begin{equation}
  c_{\ell}=|\det A_{\ell}|=\prod_{a=1}^{D}\sigma_{a,\ell},
  \label{eq:convergence}
\end{equation}
where \(\sigma_{a,\ell}\) are the singular values of \(A_\ell\). Since
\(A_\ell\) is the overlap between two orthonormal \(D\)-dimensional subspaces,
\(0\leq\sigma_{a,\ell}\leq1\). Consequently, \(c_\ell\) close to unity also
excludes an isolated poorly overlapping direction. If the path is sufficiently
fine and the same subspace is followed smoothly, these singular values remain
close to one. A sharp decrease of \(c_\ell\) indicates that the step may be too
large, or that the chosen quasihole manifold is mixing with other states.

\subsection{Berry matrix and phases}

After one closed braid the Hamiltonian has returned to its initial form, but
the parallel-transported final basis need not be equal to the initial basis.
For the oriented cycle \(\CC\), we define the remaining mismatch as the Berry
matrix
\begin{equation}
  \BB=\Psi_L^{\dagger}\Psi_0.
  \label{eq:BerryMatrix}
\end{equation}
This operational convention is used for all phases reported below. Reversing the orientation of the path reverses the sign of the quoted phase. If the path is closed only up to a known symmetry or gauge transformation, that operator should be inserted in the final overlap.  For the pinning-potential cycles considered here, the final Hamiltonian is the same as the initial one, so Eq.~\eqref{eq:BerryMatrix} is used directly.

Let \(\lambda_a\), \(a=1,\ldots,D\), be the eigenvalues of \(\BB\).  We first
map their phases to the principal interval \([0,2)\), in units of \(\pi\),
\begin{equation}
  \theta_a^{(0)}=
  \mathrm{mod}\!\left(\frac{\arg\lambda_a}{\pi},2\right)
  \in[0,2).
  \label{eq:eigenphases}
\end{equation}
Because each eigenphase is defined modulo \(2\), we also report an adjacent
branch when a principal phase lies within \(0.1\) of the upper endpoint of
\([0,2)\). Such near-cut phases are shifted by \(-2\), while all others retain
their principal values; the resulting values are denoted by \(\theta_a\).
This diagnostic adjustment depends only on the phase location relative to the
branch cut and not on the expected statistical value. As in the main text, we
define the dimensionless full-manifold mean
\begin{equation}
  \overline{\theta}_{\rm B}=\frac{1}{D}\sum_{a=1}^{D}\theta_a .
  \label{eq:mean_berry_phase}
\end{equation}
For the loop orientation in Eq.~\eqref{eq:commutator_loop} and the Berry-matrix
definition in Eq.~\eqref{eq:BerryMatrix}, the reference value used in the main
text is \(\overline{\theta}_{\rm B}=2\nu_{\rm eff}\). For a Laughlin state at
\(\nu=1/m\), this coincides with the standard Abelian quasihole full-braid
phase~\cite{Arovas1984SM}. The total phase of the manifold satisfies
\begin{equation}
  \Theta_{\mathrm{tot}}=\arg\det\BB
  =\pi D\overline{\theta}_{\rm B}
  \quad (\mathrm{mod}\;2\pi).
  \label{eq:total_phase}
\end{equation}
For the commutator loop, the numerically evaluated matrix satisfies
\(\det\BB\simeq1\), and Eq.~\eqref{eq:total_phase} consequently restricts a
consistently chosen mean to
\begin{equation}
  \overline{\theta}_{\rm B}\simeq\frac{2m}{D},
  \qquad m\in\mathbb Z.
  \label{eq:commutator_mean_constraint}
\end{equation}
This constraint fixes the allowed spacing of the mean but does not select the
integer \(m\).  The main-text result \(\overline{\theta}_{\rm B}=10/11\) for
\(D=11\) selects \(m=5\); the \(D=14\) robustness calculation below selects
\(m=4\), giving \(\overline{\theta}_{\rm B}=4/7\).

We retain the full eigenphase spectrum because \(\BB\) need not be
proportional to the identity for the finite, weakly split pinned manifolds
studied here. The branch-consistent full-manifold mean is the quantity
compared with the reference value \(2\nu_{\rm eff}\).

\subsection{Numerical checks}

At each point of the path we monitor the bandwidth of the selected
\(D\)-dimensional manifold and the gap above it,
\begin{equation}
  \delta_{\ell}=E_{D,\ell}-E_{1,\ell},
  \qquad
  \Delta_{\ell}=E_{D+1,\ell}-E_{D,\ell}.
  \label{eq:bandwidth_gap}
\end{equation}
We use the dimensionless ratio
\(r_{\ell}=\delta_{\ell}/\Delta_{\ell}\) as a measure of spectral
isolation and the overlap diagnostic
\(c_{\ell}=|\det A_{\ell}|\) defined in
Eq.~\eqref{eq:convergence} as a measure of the continuity of the
tracked subspace. A well-resolved calculation is indicated by
\(c_{\ell}\) remaining close to unity and \(r_{\ell}<1\) throughout
the path, preferably with \(r_{\ell}\ll1\). We characterize the
worst spectral separation along a braid by
\begin{equation}
  r_{\max}=\max_{\ell}\frac{\delta_{\ell}}{\Delta_{\ell}}.
  \label{eq:max_bandwidth_gap}
\end{equation}

\begin{figure}[htbp]
  \centering
  \includegraphics[width=\columnwidth]{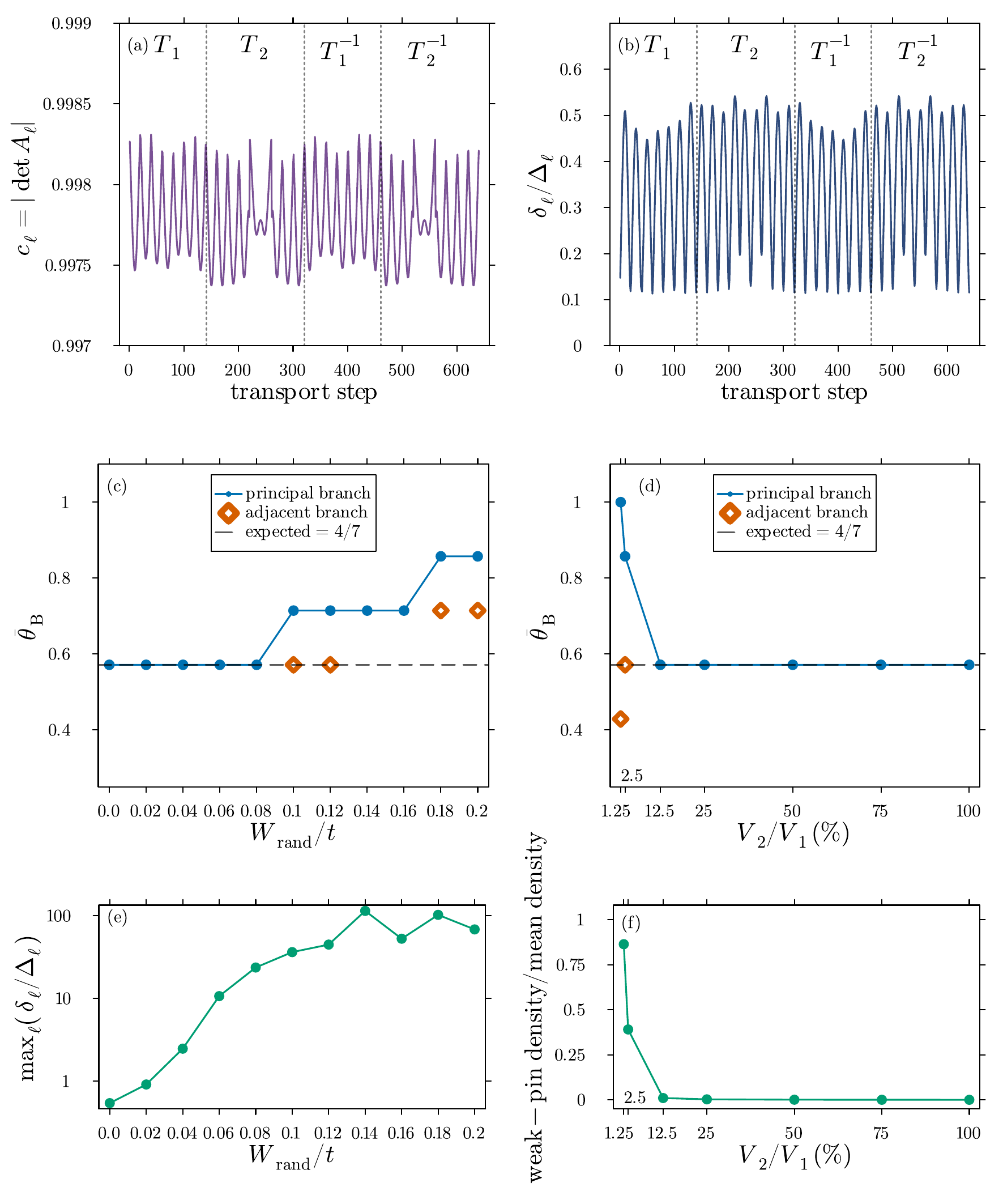}
  \caption{\label{fig:braiding_robustness}
  Robustness diagnostics for the Aharonov--Bohm-free cross-shaped braid
  of \(N=2\) bosons on a \(7\times9\) lattice with
  \(\phi=1/7\), corresponding to \(N_{\phi}=9\). Two quasiholes are
  localized by the pinning potentials, giving a \(D=14\) manifold and
  \(\nu_{\mathrm{eff}}
  =N/(N_{\phi}-N_{\mathrm{loc}})=2/7\). The plotted mean
  \(\overline{\theta}_{\rm B}\) takes the value
  \(2\nu_{\mathrm{eff}}=4/7\).
  For the clean, symmetrically pinned system
  (\(V_1=V_2=0.8t\)), panel (a) shows the overlap diagnostic
  \(c_{\ell}=|\det A_{\ell}|\), while panel (b) shows the
  instantaneous bandwidth-to-gap ratio
  \(\delta_{\ell}/\Delta_{\ell}\). The vertical dotted lines separate
  the four transport segments.
  Panels (c) and (e) show, respectively, the mean phase and
  \(\max_{\ell}(\delta_{\ell}/\Delta_{\ell})\) as a function of a random
  background potential of strength \(W_{\mathrm{rand}}/t\). Panels (d) and (f) test asymmetric pinning by fixing
  \(V_1=0.8t\) and decreasing \(V_2/V_1\); panel (f) shows the density
  at the weaker pin normalized by the spatially averaged density.
  Blue circles denote the principal phase branch, while orange diamonds correspond to the adjacent branch where shown.}
\end{figure}

The clean symmetric calculation is well conditioned. As shown in
Fig.~\ref{fig:braiding_robustness}(a), \(c_{\ell}\) remains close to
unity throughout all four segments of the braid. The ratio
\(\delta_{\ell}/\Delta_{\ell}\) in panel (b) remains below
approximately \(0.55\), so that the selected manifold remains
identifiable along the complete path. For comparison, its maximum is
below \(0.23\) for the larger five-particle system presented
in the main text, where the spectral separation is therefore still
cleaner. As a numerical null test, we also kept one pin fixed while the other traversed a path and then exactly retraced it. All bare phases were below \(10^{-13}\),
confirming numerical cancellation on a retraced path.

We next test robustness against a random background potential and unequal
pinning strengths. For the random-background test, independent site values are drawn uniformly
from \([-1,1]\), their spatial mean is subtracted, and the resulting profile
is rescaled so that \(\max_i|w_i|=W_{\mathrm{rand}}\). The same fixed
realization is rescaled for every \(W_{\mathrm{rand}}\) and
remains static throughout the braid. Panels (c) and (e) probe the response to a random background
potential. The principal-branch mean remains at the expected value
\(\overline{\theta}_{\rm B}=4/7\), equivalently \(m=4\) in
Eq.~\eqref{eq:commutator_mean_constraint}, through
\(W_{\mathrm{rand}}/t=0.08\), and changes branch for
stronger disorder. The spectral diagnostic, however, deteriorates
before the end of this phase plateau: \(r_{\max}\) grows rapidly and
exceeds unity beyond \(W_{\mathrm{rand}}/t=0.02\), eventually becoming much
larger than unity. Thus, stability of the
principal-branch phase through \(W_{\mathrm{rand}}/t=0.08\) should not be identified
with uniformly good spectral isolation over that entire range. In
particular, although the adjacent branch happens to reproduce \(4/7\)
at \(W_{\mathrm{rand}}/t=0.10\) and \(0.12\), the quasihole manifold is then poorly
separated from the remaining spectrum, and these agreements cannot
be regarded as reliable evidence for continued phase quantization.

The asymmetric-pinning test gives a consistent localization
criterion. In Fig.~\ref{fig:braiding_robustness}(d), the principal
branch yields \(\overline{\theta}_{\rm B}=4/7\) from
\(V_2/V_1=100\%\) down to and including
\(12.5\%\). Below this value, the principal-branch phase departs from \(4/7\). Although the adjacent branch gives \(4/7\) at
\(V_2/V_1=2.5\%\), panel (f) shows that the density at the weaker pin
rises rapidly once \(V_2/V_1\) is reduced below \(12.5\%\). The
weaker potential therefore no longer supports a well-localized
quasihole, and the adjacent-branch agreement at \(2.5\%\) is not
supported by the localization diagnostic.

In summary, at each transport step we diagonalize the instantaneous
Hamiltonian and align the selected low-energy subspace with that of
the preceding step using the SVD of the overlap matrix. After
completing the closed path, the eigenphases of the mismatch matrix
\(\Psi_L^{\dagger}\Psi_0\) give the geometric Berry-matrix spectrum
accumulated by the quasihole manifold. Within the parameter ranges identified by the diagnostics above, the
branch-consistent mean continues to select the same integer \(m=4\).
The overlap, spectral,
and localization diagnostics in
Fig.~\ref{fig:braiding_robustness} provide independent checks that
the extracted phase represents a well-defined braid rather than an
accidental choice of phase branch.

\section{Variational evaluation of the ansatz subspace}
\label{sec:vmc_projection_ansatz_subspace}

We evaluate the same analytic composite-boson trial-state family in two complementary ways.  For the smaller systems, the trial states are explicitly projected onto the many-body basis of the lowest Hofstadter band, and the lowest-band-projected Hamiltonian is diagonalized within their projected span.  For larger systems, VMC evaluates the overlap and Hamiltonian matrices directly in the hard-core real-space configuration basis.

Throughout this section, $|\Phi_\mu\rangle$ denotes a physical positive-flux
trial state, namely the complex conjugate of the corresponding reference state
as in Eq.~\eqref{eq:reference_to_physical_state}.

\subsection{Exact projection to the lowest Hofstadter band}
\label{subsec:exact_projected_cb}

Let $\varphi_a(r)$, with $a=1,\ldots,\Nphi$, denote the single-particle
orbitals retained in the lowest Hofstadter band.  We write a real-space
bosonic occupation state as
\begin{equation}
|\mathbf n\rangle=|n_1,\ldots,n_{N_s}\rangle,
\qquad
\sum_{r=1}^{N_s}n_r=N,
\end{equation}
and a state in the projected orbital basis as
\begin{equation}
|\mathbf m\rangle=|m_1,\ldots,m_{\Nphi}\rangle,
\qquad
\sum_{a=1}^{\Nphi}m_a=N.
\end{equation}
Introducing site and orbital creation operators related by
\begin{equation}
    a_a^\dagger=\sum_r \varphi_a(r)b_r^\dagger ,
\end{equation}
the normalized occupation states can equivalently be written as
\begin{equation}
    |\mathbf n\rangle
    =
    \frac{\displaystyle\prod_r (b_r^\dagger)^{n_r}}
         {\sqrt{\displaystyle\prod_r n_r!}}|0\rangle,
    \qquad
    |\mathbf m\rangle
    =
    \frac{\displaystyle\prod_a (a_a^\dagger)^{m_a}}
         {\sqrt{\displaystyle\prod_a m_a!}}|0\rangle .
\end{equation}
Their overlap is therefore
\begin{equation}
    \langle\mathbf m|\mathbf n\rangle
    =
    \frac{
        \langle0|
        \displaystyle\prod_a a_a^{m_a}
        \prod_r (b_r^\dagger)^{n_r}
        |0\rangle
    }{
        \sqrt{
            \displaystyle
            \prod_a m_a!
            \prod_r n_r!
        }
    } .
\end{equation}

Using
\begin{equation}
    a_a=\sum_r \varphi_a^*(r)b_r ,
\end{equation}
we introduce lists \(\{a_i\}_{i=1}^{N}\) and
\(\{r_j\}_{j=1}^{N}\), in which each orbital label \(a\) occurs \(m_a\)
times and each site label \(r\) occurs \(n_r\) times. Expanding the
product of orbital annihilation operators gives
\[
    \prod_{i=1}^{N}a_{a_i}
    =
    \sum_{s_1,\ldots,s_N}
    \left[
        \prod_{i=1}^{N}\varphi_{a_i}^{*}(s_i)
    \right]
    \prod_{i=1}^{N}b_{s_i},
\]
where the \(s_i\) are dummy site labels. The bosonic commutation
relations then give
\begin{equation}
    \langle0|
    \prod_{i=1}^{N} b_{s_i}
    \prod_{j=1}^{N} b_{r_j}^\dagger
    |0\rangle
    =
    \sum_{\sigma\in S_N}
    \prod_{i=1}^{N}
    \delta_{s_i,r_{\sigma(i)}} .
\end{equation}
Here the permutation \(\sigma\) specifies how the \(N\) annihilation
operators are paired with the \(N\) creation operators. Because the
operators are bosonic, all pairings enter with the same sign. Using the
Kronecker deltas to perform the sums over the \(s_i\) gives
\begin{equation}
    \sum_{\sigma\in S_N}
    \prod_{i=1}^{N}
    \varphi_{a_i}^*\!\left(r_{\sigma(i)}\right)
    =
    \operatorname{perm}
    \!\left[
        \varphi_{a_i}^*(r_j)
    \right]_{i,j=1}^{N},
\end{equation}
where the expression on the right is the permanent of the
$N\times N$ matrix with elements
$M_{ij}=\varphi_{a_i}^*(r_j)$. The permanent arises because all possible pairings of the bosonic
annihilation and creation operators contribute with the same sign.

Hence the many-body overlap matrix entering the band projection is
\begin{equation}
    \mathcal T_{\mathbf m\mathbf n}
    =
    \langle \mathbf m | \mathbf n \rangle
    =
    \frac{
        \operatorname{perm}
        \!\left[
            \varphi_{a_i}^{*}(r_j)
        \right]_{i,j=1}^{N}
    }{
        \sqrt{
            \displaystyle
            \prod_{a=1}^{\Nphi} m_a!
            \prod_{r=1}^{N_s} n_r!
        }
    } .
    \label{eq:projected_cb_basis_transform}
\end{equation}
Repeated orbital and site occupations therefore appear as repeated
rows and columns, respectively, of the matrix entering the permanent.

Let
\begin{equation}
C_{\mathbf n\mu}
=
\langle\mathbf n|\Phi_\mu\rangle
\end{equation}
be the matrix whose columns contain the real-space amplitudes of the raw CB states.  Since this set can be overcomplete, we first perform a singular-value decomposition of $C$ and retain only its nonzero singular directions.  This gives an orthonormal matrix $Q_{\rm CB}$ whose columns span exactly the same variational subspace as the original ansatz states. Such linear dependencies commonly arise when distinct composite-particle configurations become identical or linearly dependent after projection~\cite{MeyerLiaboeViefers2016}.  We remove them numerically by retaining only the nonzero singular directions of the coefficient matrix.

The ansatz subspace in the lowest-band many-body basis is then
\begin{equation}
\widetilde Q_{\rm CB}
=
\mathcal T Q_{\rm CB}.
\end{equation}
A second singular-value decomposition removes any additional linear dependencies generated by the band projection.  Denoting the retained orthonormal basis by $Q_{\rm proj}$, we diagonalize
\begin{equation}
H_{\rm CB}^{\rm proj}
=
Q_{\rm proj}^{\dagger}
H_{\rm proj}
Q_{\rm proj},
\label{eq:projected_cb_ritz_hamiltonian}
\end{equation}
where $H_{\rm proj}$ is the interacting Hamiltonian projected to the lowest Hofstadter band. The eigenvalues of Eq.~\eqref{eq:projected_cb_ritz_hamiltonian} are the Rayleigh--Ritz variational energies within the projected CB subspace.  This procedure is closely analogous to composite-fermion diagonalization, in which the interaction Hamiltonian is diagonalized within a restricted space of projected composite-fermion states rather than in the full many-body Hilbert space~\cite{MandalJain2002,MeyerSreejithViefers2014}.

To compare the projected CB span directly with the exact low-energy
manifold, let \(Q_{\rm ED}\) contain the lowest \(\mathcal D\)
orthonormal eigenvectors of \(H_{\rm proj}\).  We define the
basis-independent subspace fidelity
\begin{equation}
\mathcal F_{\rm sub}
=
\frac{1}{\mathcal D}
\left\|Q_{\rm proj}^{\dagger}Q_{\rm ED}\right\|_{\rm F}^{2}
=
\frac{1}{\mathcal D}\sum_{i=1}^{\mathcal D}s_i^2,
\label{eq:subspace_fidelity}
\end{equation}
where \(s_i\) are the singular values of
\(Q_{\rm proj}^{\dagger}Q_{\rm ED}\), equivalently the cosines of the
principal angles between the two subspaces.  The smallest value
\(s_{\min}\) measures their worst-aligned direction.

For all four systems investigated in the main text, \(\mathcal F_{\rm sub}>0.99969\) and
\(s_{\min}>0.99975\) as can be seen in Table~\ref{tab:subspace_fidelity}.  In particular,
\(s_{\min}^{2}>0.99951\), showing that every direction in the exact
low-energy manifold is reproduced with very little weight outside the
projected CB span.  This provides a direct subspace-level validation
that is independent of the agreement between the corresponding
energy spectra.

\begin{table}[htbp]
\caption{\label{tab:subspace_fidelity}
Direct comparison between the projected CB span and the exact
low-energy manifold.  Here
\(M_{\rm low}=\binom{N_\phi+N-1}{N}\) is the dimension of the
lowest-band many-body Hilbert space.}
\begin{ruledtabular}
\begin{tabular}{ccccc}
\((N,N_\phi)\) & \(M_{\rm low}\) & \(\mathcal D\)
& \(\mathcal F_{\rm sub}\) & \(s_{\min}\) \\
\hline
\((2,8)\)  & \(36\)  & \(20\) & \(0.999776\) & \(0.999774\) \\
\((2,10)\) & \(55\)  & \(35\) & \(0.999932\) & \(0.999886\) \\
\((3,9)\)  & \(165\) & \(30\) & \(0.999692\) & \(0.999755\) \\
\((3,10)\) & \(220\) & \(50\) & \(0.999814\) & \(0.999795\) \\
\end{tabular}
\end{ruledtabular}
\end{table}

\subsection{Monte Carlo evaluation in the hard-core configuration space}
\label{subsec:vmc_variational_matrices}

For larger systems, explicitly constructing the real-space many-body basis and the transformation in Eq.~\eqref{eq:projected_cb_basis_transform} becomes impractical.  We therefore use Monte Carlo sampling to evaluate the overlap and Hamiltonian matrices of the same analytic trial-state family directly in the hard-core real-space configuration basis, without explicit projection onto the lowest Hofstadter band.  This is the Rayleigh--Ritz method in a nonorthogonal basis, with the required matrix elements estimated by sampling configurations~\cite{Metropolis1953,FoulkesMitasNeedsRajagopal2001,NightingaleMelikAlaverdian2001,UmrigarFilippi2005,ToulouseUmrigar2007,WangGeraedtsRezayiHaldane2019}.  Composite-fermion diagonalization likewise evaluates overlap and Hamiltonian matrices by Monte Carlo sampling~\cite{MandalJain2002,BalramTokeWojsJain2015}.

Let
\begin{equation}
    \{\, |\Phi_\mu\rangle \,\}_{\mu=1}^{d}
\end{equation}
be the set of ansatz states used in a given calculation.  A hard-core configuration is denoted by
\begin{equation}
    C=(i_1,\ldots,i_N), \qquad i_1<\cdots<i_N,
\end{equation}
where the ordered list specifies the occupied lattice sites.  For each configuration, the ansatz construction gives the vector of amplitudes
\begin{equation}
    \bm{\Phi}(C)
    =\bigl(\Phi_1(C),\ldots,\Phi_d(C)\bigr)^T,
    \qquad
    \Phi_\mu(C)=\langle C|\Phi_\mu\rangle .
\end{equation}
This amplitude vector is the only ansatz-specific input needed for the Monte Carlo evaluation of the variational matrices. The Hamiltonian, boundary twists, and gauge convention are those given in the main text; here we only need the fact that, for each configuration $C$, the Hamiltonian connects it to a small number of other hard-core configurations $C'$.

For a general linear combination
\begin{equation}
    |\Psi({\bf c})\rangle=\sum_{\mu=1}^{d} c_\mu |\Phi_\mu\rangle,
\end{equation}
the variational energy is
\begin{equation}
    E({\bf c})=
    \frac{{\bf c}^\dagger H {\bf c}}{{\bf c}^\dagger G {\bf c}},
\end{equation}
with
\begin{equation}
    G_{\mu\nu}=\langle \Phi_\mu|\Phi_\nu\rangle,
    \qquad
    H_{\mu\nu}=\langle \Phi_\mu|\hat H|\Phi_\nu\rangle .
\end{equation}
The best variational states inside the ansatz subspace are obtained by solving
\begin{equation}
    H {\bf c}=E G {\bf c} .
    \label{eq:vmc_generalized_eigenproblem}
\end{equation}
Equation~\eqref{eq:vmc_generalized_eigenproblem} follows by requiring the
Rayleigh quotient to be stationary with respect to \(\mathbf c\). Its
eigenvectors give the optimal linear combinations of the CB states within
the ansatz subspace, while its eigenvalues are the corresponding Ritz
energies. Because the original states are nonorthogonal and may be linearly
dependent, the Gram matrix \(G\) need not be the identity and can contain
zero or very small eigenvalues; these directions are treated below.

The numerical task is simply to estimate the two finite matrices $G$ and $H$. To do this efficiently, we sample configurations using the nonnegative weight
\begin{equation}
    \rho(C)=\sum_{\lambda=1}^{d} |\Phi_\lambda(C)|^2,
    \qquad
    p(C)=\frac{\rho(C)}{Z},
    \qquad
    Z=\sum_C \rho(C).
    \label{eq:vmc_combined_weight}
\end{equation}

This choice is convenient because one sampled configuration contributes to all matrix elements at once.  With this sampling probability,
\begin{equation}
    G_{\mu\nu}
    =Z\,\mathbb{E}_{p}\left[
    \frac{\Phi_\mu^*(C)\Phi_\nu(C)}{\rho(C)}
    \right] .
    \label{eq:vmc_gram_expectation}
\end{equation}
Here and below,
\(\mathbb E_p[f(C)]\equiv\sum_C p(C)f(C)\)
denotes the expectation value over configurations distributed according to
\(p(C)\). The unknown constant $Z$ will cancel from the generalized eigenvalue problem, so it does not have to be determined separately.

The Hamiltonian matrix is treated in the same way.  For a given sampled configuration, we first compute the action of the Hamiltonian on each ansatz state,
\begin{equation}
    (H\Phi_\nu)(C)
    =\sum_{C'} \langle C|\hat H|C'\rangle \Phi_\nu(C') .
    \label{eq:vmc_local_hamiltonian_action}
\end{equation}
The sum runs only over configurations connected to $C$ by the Hamiltonian.  For the nearest-neighbor hard-core hopping problem used here, this list is short.  The calculations reported here contain no diagonal term; if one were
present (for instance a pinning potential), it would enter through the \(C'=C\) contribution. Then
\begin{equation}
    H_{\mu\nu}
    =Z\,\mathbb{E}_{p}\left[
    \frac{\Phi_\mu^*(C)(H\Phi_\nu)(C)}{\rho(C)}
    \right] .
    \label{eq:vmc_hamiltonian_expectation}
\end{equation}

For $N_{\rm samp}$ sampled configurations $C_s$, the matrices used in practice are therefore
\begin{align}
    \widetilde G_{\mu\nu}
    &=\frac{1}{N_{\rm samp}}\sum_{s=1}^{N_{\rm samp}}
    \frac{\Phi_\mu^*(C_s)\Phi_\nu(C_s)}{\rho(C_s)},
    \label{eq:vmc_gram_estimator}
    \\
    \widetilde H_{\mu\nu}
    &=\frac{1}{N_{\rm samp}}\sum_{s=1}^{N_{\rm samp}}
    \frac{\Phi_\mu^*(C_s)(H\Phi_\nu)(C_s)}{\rho(C_s)} .
    \label{eq:vmc_hamiltonian_estimator}
\end{align}
In the infinite-sampling limit, these estimators approach the exact real-space variational matrices divided by the common normalization factor $Z$, namely $\widetilde G\to G/Z$ and $\widetilde H\to H/Z$. Consequently, the variational energies are obtained by replacing $G,H$ with $\widetilde G,\widetilde H$ in Eq.~\eqref{eq:vmc_generalized_eigenproblem}.

The Markov chain uses moves that preserve the hard-core constraint. A move
selects one occupied site and proposes either one of its nearest neighbors or,
with a smaller fixed probability, a uniformly random lattice site. Proposals
whose target site is occupied are rejected. Both proposal kernels are
symmetric, so a valid move is accepted with probability
\begin{equation}
    A(C\to C')=
    \min\left(1,\frac{\rho(C')}{\rho(C)}\right).
\end{equation}

In the actual implementation, local moves can be supplemented by occasional longer moves in order to improve the sampling of configuration space. All proposal types used here are chosen symmetrically, so the acceptance probability above applies without an additional proposal-ratio factor. The sampling weight is real and nonnegative, so no complex or signed weight is sampled; the complex phases of the ansatz states and of the hopping matrix elements enter only through the estimators in Eqs.~\eqref{eq:vmc_gram_estimator} and \eqref{eq:vmc_hamiltonian_estimator}.

After sampling, the estimated matrices are symmetrized,
\begin{equation}
    \widetilde G\leftarrow \frac{\widetilde G+\widetilde G^\dagger}{2},
    \qquad
    \widetilde H\leftarrow \frac{\widetilde H+\widetilde H^\dagger}{2} .
\end{equation}

Since the ansatz construction may generate more states than are linearly
independent, we diagonalize the Hermitian Gram matrix,
\begin{equation}
    \widetilde G=U\Lambda U^\dagger,
\end{equation}
where $\Lambda=\operatorname{diag}(\lambda_1,\lambda_2,\ldots)$
contains the Gram-matrix eigenvalues. Directions whose eigenvalues are
numerically negligible are discarded. In the retained subspace all
$\lambda_i$ are positive, and we define
\begin{equation}
    \Lambda^{-1/2}
    =
    \operatorname{diag}
    \left(
        \lambda_1^{-1/2},
        \lambda_2^{-1/2},
        \ldots
    \right).
\end{equation}
Writing the generalized-eigenproblem coefficients as
\begin{equation}
    {\bf c}=U\Lambda^{-1/2}{\bf v},
\end{equation}
one finds
\begin{align}
    \widetilde H U\Lambda^{-1/2}{\bf v}
    &=
    E\,\widetilde G U\Lambda^{-1/2}{\bf v}
    \notag\\
    &=
    E\,U\Lambda^{1/2}{\bf v}.
\end{align}
Multiplication from the left by $\Lambda^{-1/2}U^\dagger$ therefore
gives the equivalent ordinary Hermitian eigenvalue problem
\begin{equation}
    \Lambda^{-1/2}U^\dagger
    \widetilde H U\Lambda^{-1/2}\,{\bf v}
    =
    E\,{\bf v}.
    \label{eq:vmc_orthonormalized_eigenproblem}
\end{equation}
Equivalently, with $X=U\Lambda^{-1/2}$ one has
$X^\dagger\widetilde G X=I$, so the retained nonorthogonal ansatz
states have been transformed to an orthonormal basis.

This step is also useful diagnostically: the number of retained Gram-matrix
eigenvalues gives the numerical rank of the ansatz subspace. For the smaller
systems, the corresponding blockwise analysis of the raw translated states
gives the orbit-resolved ranks reported in
Table~\ref{tab:orbit-rank-test}.

Finally, we estimate the statistical uncertainties by dividing the Markov
chain into \(B=n_{\rm bin}\) consecutive bins.  Let
\(\widetilde G^{(b)}\) and \(\widetilde H^{(b)}\) denote the matrices
averaged within bin \(b\).  The central energies are obtained from the
matrices averaged over all bins,
\begin{equation}
    \widetilde G=\frac{1}{B}\sum_{b=1}^{B}\widetilde G^{(b)},
    \qquad
    \widetilde H=\frac{1}{B}\sum_{b=1}^{B}\widetilde H^{(b)}.
\end{equation}
To propagate the statistical fluctuations through the nonlinear generalized
eigenvalue problem, we omit each bin in turn and repeat the complete
diagonalization, i.e., we use a leave-one-bin-out jackknife,
treating the bin averages as the jackknife observations~\cite{Efron1982Jackknife}.
For the replica in which bin \(b\) is omitted,
we form
\begin{equation}
    \widetilde G^{(-b)}
    =\frac{1}{B-1}\sum_{b'\ne b}\widetilde G^{(b')},
    \qquad
    \widetilde H^{(-b)}
    =\frac{1}{B-1}\sum_{b'\ne b}\widetilde H^{(b')},
\end{equation}
and solve
\begin{equation}
    \widetilde H^{(-b)}{\bf c}
    =E_n^{(-b)}\widetilde G^{(-b)}{\bf c}.
\end{equation}
The quoted standard error of level \(n\) is
\begin{equation}
    \sigma_{E_n}
    =
    \left[
    \frac{B-1}{B}
    \sum_{b=1}^{B}
    \left(E_n^{(-b)}-\overline E_n^{(-)}\right)^2
    \right]^{1/2},
    \qquad
    \overline E_n^{(-)}
    =\frac{1}{B}\sum_{b=1}^{B}E_n^{(-b)} .
    \label{eq:vmc_jackknife_error}
\end{equation}
The Gram-matrix rank is recomputed for every jackknife replica and is required
to remain stable.

For both calculations reported in the main text, the first
\(2.0\times10^4\) Metropolis update attempts were discarded for
thermalization, and successive recorded configurations were separated by
\(100\) further update attempts. The \((N,N_\phi)=(5,11)\) calculation
used \(1500\) configurations in each of \(150\) bins, giving
\(2.25\times10^5\) sampled configurations in total. The corresponding
numbers for \((N,N_\phi)=(6,14)\) were \(1000\) configurations per bin and
\(150\) bins, giving \(1.50\times10^5\) configurations. A relative
Gram-eigenvalue cutoff of \(10^{-8}\) was used. All \(150\) jackknife
replicas retained the predicted ranks \(\mathcal D=11\) and \(49\),
respectively.

For comparison with the lowest-band-projected results, we define
\(\Delta=E^{\rm proj}_{\mathcal D+1}-E^{\rm proj}_{\mathcal D}\) as the
projected gap above the manifold. The total VMC energy spreads
\(E^{\rm VMC}_{\mathcal D}-E^{\rm VMC}_{1}\) are \(0.0517\Delta\) and
\(0.1126\Delta\) for \((5,11)\) and \((6,14)\), respectively. After
ordering the levels by energy, every VMC level lies within \(0.1211\Delta\)
of its lowest-band-projected counterpart, while the largest jackknife
standard error is \(0.0059\Delta\). Here the VMC values are variational
energies of the hard-core real-space Hamiltonian, whereas
\(E_n^{\rm proj}\) denotes the corresponding lowest-band-projected energy.
This is therefore a robustness comparison across related Hamiltonians, not a
variational-error estimate.

In summary, the calculation uses VMC only as a controlled numerical summation method for the real-space variational matrices of the finite CB trial-state family; no explicit lowest-band projection is performed.  Once the matrices $\widetilde G$ and $\widetilde H$ have been estimated, the remaining step is a standard generalized Rayleigh--Ritz diagonalization in the retained nonorthogonal trial-state basis.

\bibliographystyle{apsrev4-2}
\bibliography{supplemental_material_references}